\documentclass[10pt,letterpaper]{article}
\usepackage{geometry}
\usepackage{epsfig}
\usepackage{amsmath,amssymb,slashed,bbm}
\usepackage{setspace}
\csname@addtoreset\endcsname{equation}{section}

\makeatletter
\pdfoutput=1

\usepackage[T1]{fontenc}
\usepackage{ae}
\usepackage{aecompl}

\usepackage[english]{babel}
\usepackage[sf,bf,small]{titlesec}
\usepackage[small,sf,bf,hang]{caption}

\usepackage[multiple]{footmisc}
\long\def\symbolfootnote[#1]#2{\begingroup%
\def\thefootnote{\fnsymbol{footnote}}\footnote[#1]{#2}\endgroup}

\usepackage{titletoc}
\dottedcontents{section}[6mm]{\smallskip }{6mm}{0pc} 
\dottedcontents{subsection}[8mm]{\itshape\small }{6mm}{0pc}
\dottedcontents{subsubsection}[16mm]{\itshape\small }{10mm}{0pc}
\def\tableofcontents{\subsection*{\contentsname}\vspace{-2mm}\@starttoc{toc}}

\usepackage[nosort]{cite} 

\usepackage[parsep]{collref}

\usepackage{feynmp}  
\DeclareGraphicsExtensions{.pdf}  % include pdf files by default
\renewcommand{\bar}[1]{\overline{#1}}
\def \bea  {\begin{eqnarray}}
\def \eea  {\end{eqnarray}}

\newcommand{\nn}{\nonumber}

\newcommand{\ns} \normalsize

\newcommand{\3}{$AdS_3\times S^3\times S^3\times S^1$}

\AtBeginDocument{
  
}

\makeatother

\begin{document}
%\selectlanguage{english}
\begin{flushright}
MIFPA-12-26\bigskip\bigskip
\par\end{flushright}

\begin{center}
\textsf{\textbf{\Large Classical integrability and quantum aspects of the $AdS_3\times S^3\times S^3\times S^1$ superstring\smallskip\smallskip }}\\
\textsf{\textbf{\Large  }}
\par\end{center}{\Large \par}

\begin{singlespace}
\begin{center}
Per Sundin$^{1}$ and  Linus Wulff$^{2}$ \bigskip \\

{\small $^{1}$}\emph{\small{} The Laboratory for Quantum Gravity \& Strings}\\
\emph{\small Department of Mathematics and Applied Mathematics, }\\
\emph{\small University of Cape Town,}\\
\emph{\small Private Bag, Rondebosch, 7700, South Africa}{\small }\\
\emph{\small nidnus.rep@gmail.com}\vspace{0.2cm}

{\small $^{2}$}\emph{\small{} George P. \& Cynthia Woods Mitchell Institute for Fundamental Physics and Astronomy,}\\
\emph{\small Texas A\&M University, College Station, }\\
\emph{\small TX 77843, USA}\\
\emph{\small linus@physics.tamu.edu }{\small \bigskip }\emph{\small }\\
\par\end{center}{\small \par}
\end{singlespace}

%\begin{center}
%Draft printed \bigskip  \today , from file \jobname.
%\par\end{center}

\subsection*{\hspace{9mm}Abstract}
\begin{quote}
In this paper we continue the investigation of aspects of integrability of the type IIA \3 and $AdS_3\times S^3\times T^4$ superstrings. By constructing a one parameter family of flat connections we prove that the Green-Schwarz string is classically integrable, at least to quadratic order in fermions, without fixing the kappa-symmetry. We then compare the quantum dispersion relation, fixed by integrability up to an unknown interpolating function $h(\lambda)$, to explicit one-loop calculations on the string worldsheet. For \3 the spectrum contains heavy, as well as light and massless modes, and we find that the one-loop contribution differs depending on how we treat these modes showing that similar regularization ambiguities as appeared in AdS$_4$ / CFT$_3$ occur also here. 
  \bigskip  \thispagestyle{empty} 
\end{quote}
%%
%% This file contains descriptions of feynman diagrams for use with the package feynMP
%% After running the main file, run "mpost diagrams" once to generate these
%%
%%
%% These are for propagators paper with Per. 
%% The version in this folder edited 18 May 2011 to add unlabled ones for introduction. 
%%

\newsavebox{\feynmanrules}
\sbox{\feynmanrules}{
\begin{fmffile}{diagrams} % I can't seem to make this work using any path but the same one as the document

%%%%%%%%%%%%%%%%
%%  SETTINGS

\fmfset{thin}{0.6pt}  % was 0.7 until v24
%\fmfset{wiggly_len}{5mm}
\fmfset{dash_len}{4pt}
\fmfset{dot_size}{1thick}
\fmfset{arrow_len}{6pt} % you can't use em here, mpost doesn't know what it will be.
%\fmfset{curly_len}{2.5mm}
%\setlength{\unitlength}{1em} % default is =1pt, maybe that's sensible. 72pt = 1in

%%%%%%%%%%%%%%%%%%%
%% HEAVY-MODE DECAY PROCESSES

\begin{fmfgraph*}(100,36)
\fmfkeep{decay1}
\fmfleft{v1}
\fmfright{o1,o2}
\fmfdot{v2}
\fmf{dashes_arrow,label=\small{$y_1$}}{v1,v2}
\fmf{fermion}{v2,o2}
\fmf{fermion}{v2,o1}
\fmflabel{\small{$\chi_\pm^{(3)}$}}{o1}
\fmflabel{\small{$\chi_\pm^{(2)}$}}{o2}
\end{fmfgraph*}

\begin{fmfgraph*}(100,36)
\fmfkeep{decay2}
\fmfleft{v1}
\fmfright{o1,o2}
\fmfdot{v2}
\fmf{fermion,label=\small{$\chi_\pm^{(1)}$}}{v1,v2}
\fmf{dashes_arrow}{v2,o2}
\fmf{fermion}{v2,o1}
\fmflabel{\small{$\chi_\pm^{(3)}$}}{o1}
\fmflabel{\small{$y_2$}}{o2}
\end{fmfgraph*}

\begin{fmfgraph*}(100,36)
\fmfkeep{decay3}
\fmfleft{v1}
\fmfright{o1,o2}
\fmfdot{v2}
\fmf{fermion,label=\small{$\chi_\pm^{(1)}$}}{v1,v2}
\fmf{dashes_arrow}{v2,o2}
\fmf{fermion}{v2,o1}
\fmflabel{\small{$\chi_\pm^{(2)}$}}{o1}
\fmflabel{\small{$y_3$}}{o2}
\end{fmfgraph*}

%%%%%%%%%%%%%%%%%%%%

%% TREE LEVEL SCATTERING

\begin{fmfgraph*}(50,25)
\fmfkeep{S-t-chan}
\fmfbottom{i1,d1,o1}
\fmftop{i2,d2,o2}
\fmf{fermion}{i1,v1,o1}
\fmf{fermion}{i2,v2,o2}
\fmf{plain,tension=0}{v1,v2}
\fmflabel{$(t)$}{v1}
\fmflabel{$2$}{i1}
\fmflabel{$1$}{i2}
\fmflabel{$4$}{o1}
\fmflabel{$3$}{o2}
\end{fmfgraph*}

\begin{fmfgraph*}(50,25)
\fmfkeep{S-u-chan}
\fmfbottom{i1,d1,o1}
\fmftop{i2,d2,o2}
\fmf{fermion,tension=2}{i1,v1}
\fmf{phantom,tension=1.5}{v1,o1}
\fmf{plain}{v1,v3}
\fmf{fermion}{v3,o2}
\fmf{fermion,tension=2}{i2,v2}
\fmf{phantom,tension=1.5}{v2,o2}
\fmf{plain}{v2,v3}
\fmf{fermion}{v3,o1}
\fmf{plain,tension=-0.5}{v1,v2}
\fmflabel{$(u)$}{v1}
\fmflabel{$2$}{i1}
\fmflabel{$1$}{i2}
\fmflabel{$4$}{o1}
\fmflabel{$3$}{o2}
\end{fmfgraph*}

\begin{fmfgraph*}(50,25)
\fmfkeep{T-t-chan}
\fmfbottom{i1,d1,o1}
\fmftop{i2,d2,o2}
\fmf{fermion}{o1,v1,i1}
\fmf{fermion}{i2,v2,o2}
\fmf{plain,tension=0}{v1,v2}
\fmflabel{$(t)$}{v1}
\fmflabel{$2$}{i1}
\fmflabel{$1$}{i2}
\fmflabel{$4$}{o1}
\fmflabel{$3$}{o2}
\end{fmfgraph*}

\begin{fmfgraph*}(50,25)
\fmfkeep{T-s-chan}
\fmfleft{i1,i2}
\fmfright{o1,o2}
\fmf{fermion}{i2,v1,i1}
\fmf{fermion}{o1,v2,o2}
\fmf{plain,label=$(s)$,label.dist=16}{v1,v2}
\fmflabel{$2$}{i1}
\fmflabel{$1$}{i2}
\fmflabel{$4$}{o1}
\fmflabel{$3$}{o2}
\end{fmfgraph*}

\begin{fmfgraph*}(50,25)
\fmfkeep{R-u-chan}
\fmfbottom{i1,d1,o1}
\fmftop{i2,d2,o2}
\fmf{fermion,tension=2}{i1,v1}
\fmf{phantom,tension=1.5}{v1,o1}
\fmf{plain}{v1,v3}
\fmf{fermion}{o2,v3}
\fmf{fermion,tension=2}{i2,v2}
\fmf{phantom,tension=1.5}{v2,o2}
\fmf{plain}{v2,v3}
\fmf{fermion}{v3,o1}
\fmf{plain,tension=-0.5}{v1,v2}
\fmflabel{$(t)$}{v1}
\fmflabel{$2$}{i1}
\fmflabel{$1$}{i2}
\fmflabel{$4$}{o1}
\fmflabel{$3$}{o2}
\end{fmfgraph*}

\begin{fmfgraph*}(50,25)
\fmfkeep{R-s-chan}
\fmfleft{i1,i2}
\fmfright{o1,o2}
\fmf{fermion}{i2,v1}
\fmf{fermion}{i1,v1}
\fmf{fermion}{v2,o1}
\fmf{fermion}{o2,v2}
\fmf{plain,label=$(s)$,label.dist=16}{v1,v2}
\fmflabel{$2$}{i1}
\fmflabel{$1$}{i2}
\fmflabel{$4$}{o1}
\fmflabel{$3$}{o2}
\end{fmfgraph*}

\begin{fmfgraph*}(50,25)
\fmfkeep{S-c}
\fmfbottom{i1,o1}
\fmftop{i2,o2}
\fmf{fermion}{i1,v1,o1}
\fmf{fermion}{i2,v1,o2}
%\fmf{plain,label=$(s)$}{v1}
%\fmf{fermion,tension=0}{v1,v2}
\fmf{phantom,label=$(c)$}{i1,o1}
\fmflabel{$(t)$}{i2}
\fmflabel{$2$}{i1}
\fmflabel{$1$}{i2}
\fmflabel{$4$}{o1}
\fmflabel{$3$}{o2}
\end{fmfgraph*}

\begin{fmfgraph*}(50,25)
\fmfkeep{T-c}
\fmfbottom{i1,o1}
\fmftop{i2,o2}
\fmf{fermion}{i2,v1,i1}
\fmf{fermion}{o1,v1,o2}
\fmf{phantom,label=$(c)$}{i1,o1}
%\fmf{fermion,tension=0}{v1,v2}
%\fmflabel{$(t)$}{v1}
\fmflabel{$2$}{i1}
\fmflabel{$1$}{i2}
\fmflabel{$4$}{o1}
\fmflabel{$3$}{o2}
\end{fmfgraph*}

\begin{fmfgraph*}(50,25)
\fmfkeep{rrr-c}
\fmfbottom{i1,o1}
\fmftop{i2,o2}
\fmf{fermion}{i2,v1,i1}
\fmf{fermion}{o1,v1,o2}
\fmf{phantom,label=$(c)$}{i1,o1}
%\fmf{fermion,tension=0}{v1,v2}
%\fmflabel{$(t)$}{v1}
\fmflabel{$2$}{i1}
\fmflabel{$1$}{i2}
\fmflabel{$3$}{o1}
\fmflabel{$4$}{o2}
\end{fmfgraph*}

\begin{fmfgraph*}(50,25)
\fmfkeep{oldR-c}
\fmfbottom{i1,o1}
\fmftop{i2,o2}
\fmf{fermion}{i1,v1,o2}
\fmf{fermion}{i2,v1,o1}
\fmf{phantom,label=$(c)$}{i1,o1}
%\fmf{fermion,tension=0}{v1,v2}
%\fmflabel{$(t)$}{v1}
\fmflabel{$2$}{i1}
\fmflabel{$1$}{i2}
\fmflabel{$4$}{o1}
\fmflabel{$3$}{o2}
\end{fmfgraph*}

%%%%%%%%%%%%%%%%%%%%
%% TADPOLE GRAPHS  

 \begin{fmfgraph*}(110,62)
	\fmfkeep{tadpoleaaa}
% Note that the size is given in normal parentheses
% instead of curly brackets.
% Define external vertices from bottom to top
 \fmfleft{l}
    \fmf{photon}{l,i}
    \fmf{fermion}{f1,i,f2}
    \fmfright{f1,f2}

   \end{fmfgraph*}

\begin{fmfgraph*}(72,25)
\fmfkeep{tadpole-lightlight}
\fmfleft{in,p1}
\fmfright{out,p2}
\fmfdot{c}
\fmf{dbl_dashes,label=\small{$y_1(p)$}}{in,c}
\fmf{dbl_dashes}{c,out}
\fmf{plain_arrow,right, tension=0.8, label=\small{$y_k\text{ or }\chi^k_\pm$}}{c,c}
\fmf{phantom, tension=0.2}{p1,p2}
\end{fmfgraph*}

\begin{fmfgraph*}(72,25)
\fmfkeep{tadpole-heavyheavy}
\fmfleft{in,p1}
\fmfright{out,p2}
\fmfdot{c}
\fmf{dbl_dashes,label=\small{$y_i(p)$}}{in,c}
\fmf{dbl_dashes}{c,out}
\fmf{plain_arrow,right, tension=0.8, label=\small{$y_1\text{ or }\chi^1_\pm \text{ with }\alpha\Lambda$}}{c,c}
\fmf{phantom, tension=0.2}{p1,p2}
\end{fmfgraph*}

%\begin{fmfgraph*}(72,25)
%\fmfkeep{tadpole-heavy-less}
%\fmfleft{in,p1}
%\fmfright{out,p2}
%\fmfdot{c}
%\fmf{dashes_arrow,label=\small{$ \omega_\alpha (p)$}}{in,c}
%\fmf{dashes_arrow}{c,out}
%\fmf{dbl_plain,right, tension=0.8, label=\small{ \  }}{c,c}
%\fmf{phantom, tension=0.2}{p1,p2}
%\end{fmfgraph*}

%%% y ext

\begin{fmfgraph*}(72,25)
\fmfkeep{tad-nolabel}
\fmfset{dash_len}{6pt} % this seems to be a local change
\fmfleft{in,p1}
\fmfright{out,p2}
\fmfdot{c}
\fmf{dashes_arrow,label=\small{$ y_i $}}{in,c}
\fmf{dashes_arrow}{c,out}
\fmf{plain_arrow,right, tension=0.8}{c,c}
\fmf{phantom, tension=0.2}{p1,p2}
\end{fmfgraph*}

\begin{fmfgraph*}(72,25)
\fmfkeep{tadpole-yy}
\fmfset{dash_len}{6pt} % this seems to be a local change
\fmfleft{in,p1}
\fmfright{out,p2}
\fmfdot{c}
\fmf{dbl_dashes,label=\small{$ y $}}{in,c}
\fmf{dbl_dashes}{c,out}
\fmf{plain_arrow,right, tension=0.8, label=\small{$\omega_\alpha\text{ or }\psi^a$}}{c,c}
\fmf{phantom, tension=0.2}{p1,p2}
\end{fmfgraph*}

\begin{fmfgraph*}(72,25)
\fmfkeep{tadpole-y4}
\fmfset{dash_len}{6pt} % this seems to be a local change
\fmfleft{in,p1}
\fmfright{out,p2}
\fmfdot{c}
\fmf{dashes_arrow,label=\small{$ y_4(p) $}}{in,c}
\fmf{dashes_arrow}{c,out}
\fmf{plain_arrow,right, tension=0.8, label=\small{$y_k$}}{c,c}
\fmf{phantom, tension=0.2}{p1,p2}
\end{fmfgraph*}

\begin{fmfgraph*}(72,25)
\fmfkeep{tadpole-y2bos}
\fmfset{dash_len}{6pt} % this seems to be a local change
\fmfleft{in,p1}
\fmfright{out,p2}
\fmfdot{c}
\fmf{dashes_arrow,label=\small{$ y_i(p) $}}{in,c}
\fmf{dashes_arrow}{c,out}
\fmf{plain_arrow,right, tension=0.8, label=\small{$y$}}{c,c}
\fmf{phantom, tension=0.2}{p1,p2}
\end{fmfgraph*}

\begin{fmfgraph*}(72,25)
\fmfkeep{tadpole-y2ferm}
\fmfset{dash_len}{6pt} % this seems to be a local change
\fmfleft{in,p1}
\fmfright{out,p2}
\fmfdot{c}
\fmf{dashes_arrow,label=\small{$ y_i(p) $}}{in,c}
\fmf{dashes_arrow}{c,out}
\fmf{plain_arrow,right, tension=0.8, label=\small{$\chi_\pm$}}{c,c}
\fmf{phantom, tension=0.2}{p1,p2}
\end{fmfgraph*}
%%% lollipop

\begin{fmfgraph*}(72,36)
\fmfkeep{lollipop}

\fmfstraight
\fmfleft{in,i1,i2}
\fmfright{out,o1,o2}
\fmfdot{bot,mid}

\fmf{dashes_arrow}{in,bot} %  removed ,label=\small{$\omega_\alpha (p)$}
\fmf{dashes_arrow}{bot,out}

\fmf{phantom}{i2,top}
\fmf{phantom}{top,o2}
\fmffreeze

\fmf{phantom}{i1,mid}
\fmf{phantom}{mid,o1}

\fmf{plain_arrow, tension=2}{bot,mid}

\fmf{plain_arrow,right, tension=0.8}{mid,top}
\fmf{plain,right, tension=0.8}{top,mid}
\end{fmfgraph*}

\begin{fmfgraph*}(72,36)

\fmfkeep{lollF}

\fmfstraight
\fmfleft{in,i1,i2}
\fmfright{out,o1,o2}
\fmfdot{bot,mid}

\fmf{dashes_arrow,label=\small{$y_2(p)$}}{in,bot}
\fmf{dashes_arrow}{bot,out}

\fmf{phantom}{i2,top}
\fmf{phantom}{top,o2}
\fmffreeze

\fmf{phantom}{i1,mid}
\fmf{phantom}{mid,o1}

\fmf{plain_arrow, tension=2}{bot,mid}

\fmf{plain_arrow,right, tension=0.8}{mid,top}
\fmf{plain,right, tension=0.8,label=\small{$\chi_\pm$}}{top,mid}
\end{fmfgraph*}

\begin{fmfgraph*}(72,36)
\fmfkeep{lollB}

\fmfstraight
\fmfleft{in,i1,i2}
\fmfright{out,o1,o2}
\fmfdot{bot,mid}

\fmf{dashes_arrow,label=\small{$y_2(p)$}}{in,bot} 
\fmf{dashes_arrow}{bot,out}

\fmf{phantom}{i2,top}
\fmf{phantom}{top,o2}
\fmffreeze

\fmf{phantom}{i1,mid}
\fmf{phantom}{mid,o1}

\fmf{plain_arrow, tension=2}{bot,mid}

\fmf{plain_arrow,right, tension=0.8}{mid,top}
\fmf{plain,right, tension=0.8,label=\small{$y$}}{top,mid}
\end{fmfgraph*}

%%%%%%%%%%%%%%%%%%%%
%% BUBBLE GRAPHS  

\begin{fmfgraph*}(100,36)
\fmfkeep{bubble}
\fmfleft{in}
\fmfright{out}
\fmfdot{v1}
\fmfdot{v2}
\fmf{dashes_arrow,label=\small{$\omega_\alpha (p)$}}{in,v1}
\fmf{dashes_arrow}{v2,out}
\fmf{plain_arrow,left,tension=0.6,label=\small{$\omega_\alpha (k)$}}{v1,v2}
\fmf{dbl_plain,right,tension=0.6,label=\small{$y(q)$}}{v1,v2}
\end{fmfgraph*}

%\begin{fmfgraph*}(72,25)
%\fmfkeep{tad-nolabel}
%\fmfset{dash_len}{6pt} % this seems to be a local change
%\fmfleft{in,p1}
%\fmfright{out,p2}
%\fmfdot{c}
%\fmf{dbl_dashes,label=\small{$ y_i $}}{in,c}
%\fmf{dbl_dashes}{c,out}
%\fmf{plain_arrow,right, tension=0.8}{c,c}
%\fmf{phantom, tension=0.2}{p1,p2}
%\end{fmfgraph*}

\begin{fmfgraph*}(100,36)
\fmfkeep{bubble-nolabel}
\fmfleft{in}
\fmfright{out}
\fmfdot{v1}
\fmfdot{v2}
\fmf{dashes_arrow,label=\small{$y_i$}}{in,v1}
\fmf{dashes_arrow}{v2,out}
\fmf{plain,left,tension=0.6}{v1,v2}
\fmf{plain,right,tension=0.6}{v1,v2}
%\fmf{plain_arrow,left,tension=0.6}{v1,v2}
%\fmf{plain_arrow,right,tension=0.6}{v1,v2}
\end{fmfgraph*}

\begin{fmfgraph*}(100,36)
\fmfkeep{bubble-y1}
\fmfleft{in}
\fmfright{out}
\fmfdot{v1}
\fmfdot{v2}
\fmf{dbl_dashes,label=\small{$y_1(p)$}}{in,v1}
\fmf{dbl_dashes}{v2,out}
\fmf{plain_arrow,left,tension=0.6,label=\small{$\chi_\pm^{(2)} $}}{v1,v2}
\fmf{plain_arrow,right,tension=0.6,label=\small{$\chi_\pm^{(3)}$}}{v1,v2}
\end{fmfgraph*}

\begin{fmfgraph*}(100,36)
\fmfkeep{bubble-yy}
\fmfset{dash_len}{6pt} % this seems to be a local change
\fmfleft{in}
\fmfright{out}
\fmfdot{v1}
\fmfdot{v2}
\fmf{dashes_arrow,label=\small{$ y(p) $}}{in,v1}
\fmf{dashes_arrow}{v2,out}
\fmf{plain_arrow,left,tension=0.6,label=\small{$\omega_\alpha (k)$}}{v1,v2}
\fmf{plain_arrow,left,tension=0.6,label=\small{$\omega_\alpha (q)$}}{v2,v1}
\end{fmfgraph*}

\begin{fmfgraph*}(100,36)
\fmfkeep{bubble-y2-chi24}
\fmfleft{in}
\fmfright{out}
\fmfdot{v1}
\fmfdot{v2}
\fmf{dashes_arrow,label=\small{$y_2 (p)$}}{in,v1}
\fmf{dashes_arrow}{v2,out}
\fmf{plain_arrow,left,tension=0.6,label=\small{$\chi_\pm^{(2)}$}}{v1,v2}
\fmf{plain_arrow,right,tension=0.6,label=\small{$\chi_\pm^{(4)}$}}{v1,v2}
\end{fmfgraph*}

\begin{fmfgraph*}(100,36)
\fmfkeep{bubble-y2-chi13}
\fmfleft{in}
\fmfright{out}
\fmfdot{v1}
\fmfdot{v2}
\fmf{dashes_arrow,label=\small{$y_2 (p)$}}{in,v1}
\fmf{dashes_arrow}{v2,out}
\fmf{plain_arrow,left,tension=0.6,label=\small{$\chi_\pm^{(1)}$}}{v1,v2}
\fmf{plain_arrow,right,tension=0.6,label=\small{$\chi_\pm^{(3)}$}}{v1,v2}
\end{fmfgraph*}

\begin{fmfgraph*}(100,36)
\fmfkeep{bubble-y2-y24}
\fmfleft{in}
\fmfright{out}
\fmfdot{v1}
\fmfdot{v2}
\fmf{dashes_arrow,label=\small{$y_2 (p)$}}{in,v1}
\fmf{dashes_arrow}{v2,out}
\fmf{plain_arrow,left,tension=0.6,label=\small{$y_2$}}{v1,v2}
\fmf{plain_arrow,right,tension=0.6,label=\small{$y_4$}}{v1,v2}
\end{fmfgraph*}

\begin{fmfgraph*}(100,36)
\fmfkeep{bubble-y3-chi34}
\fmfleft{in}
\fmfright{out}
\fmfdot{v1}
\fmfdot{v2}
\fmf{dashes_arrow,label=\small{$y_3 (p)$}}{in,v1}
\fmf{dashes_arrow}{v2,out}
\fmf{plain_arrow,left,tension=0.6,label=\small{$\chi_\pm^{(3)}$}}{v1,v2}
\fmf{plain_arrow,right,tension=0.6,label=\small{$\chi_\pm^{(4)}$}}{v1,v2}
\end{fmfgraph*}

\begin{fmfgraph*}(100,36)
\fmfkeep{bubble-y3-chi12}
\fmfleft{in}
\fmfright{out}
\fmfdot{v1}
\fmfdot{v2}
\fmf{dashes_arrow,label=\small{$y_3 (p)$}}{in,v1}
\fmf{dashes_arrow}{v2,out}
\fmf{plain_arrow,left,tension=0.6,label=\small{$\chi_\pm^{(1)}$}}{v1,v2}
\fmf{plain_arrow,right,tension=0.6,label=\small{$\chi_\pm^{(2)}$}}{v1,v2}
\end{fmfgraph*}

\begin{fmfgraph*}(100,36)
\fmfkeep{bubble-y3-y34}
\fmfleft{in}
\fmfright{out}
\fmfdot{v1}
\fmfdot{v2}
\fmf{dashes_arrow,label=\small{$y_3 (p)$}}{in,v1}
\fmf{dashes_arrow}{v2,out}
\fmf{plain_arrow,left,tension=0.6,label=\small{$y_3$}}{v1,v2}
\fmf{plain_arrow,right,tension=0.6,label=\small{$y_4$}}{v1,v2}
\end{fmfgraph*}

\begin{fmfgraph*}(100,36)
\fmfkeep{bubble-y4-chi23}
\fmfleft{in}
\fmfright{out}
\fmfdot{v1}
\fmfdot{v2}
\fmf{dashes_arrow,label=\small{$y_4 (p)$}}{in,v1}
\fmf{dashes_arrow}{v2,out}
\fmf{plain_arrow,left,tension=0.6,label=\small{$\chi_\pm^{(i)}$}}{v1,v2}
\fmf{plain_arrow,right,tension=0.6,label=\small{$\chi_\pm^{(i)}$}}{v1,v2}
\end{fmfgraph*}

\begin{fmfgraph*}(100,36)
\fmfkeep{bubble-y4-y23}
\fmfleft{in}
\fmfright{out}
\fmfdot{v1}
\fmfdot{v2}
\fmf{dashes_arrow,label=\small{$y_4 (p)$}}{in,v1}
\fmf{dashes_arrow}{v2,out}
\fmf{plain_arrow,left,tension=0.6,label=\small{$y_i$}}{v1,v2}
\fmf{plain_arrow,right,tension=0.6,label=\small{$y_i$}}{v1,v2}
\end{fmfgraph*}
%%%%%%%%%%%%%%%
%% THE END 

\end{fmffile}
}

\newpage
\setcounter{page}{1}
\tableofcontents{}

\newpage
\section{Introduction}
In \cite{Babichenko:2009dk} an analysis of the integrable structures of $AdS_3 / CFT_2$ was initiated. On the gravity side of the duality we have either $AdS_3\times S^3\times T^4$ or $AdS_3\times S^3\times S^3 \times S^1$, supported by pure RR-flux. For the first background the dual $CFT_2$ should be a two-dimensional sigma model on a moduli space built out of $Q_1$ instantons in a $U(Q_5)$ gauge theory on $T^4$. This is somewhat natural since $AdS_3\times S^3\times T^4$ arises as the near horizon limit of $Q_1/Q_5$ intersecting $D1/D5$ branes, \cite{Elitzur:1998mm,Gauntlett:1998kc,Cowdall:1998bu,Boonstra:1998yu,Papadopoulos:1999tw,Giveon:2003ku,deBoer:1999rh,David:2008yk,David:2010yg,Maldacena:2000hw,Maldacena:2000kv,Maldacena:2001km,Gava:2002xb,Lunin:2002fw,Gomis:2002qi,Sommovigo:2003kd,Sadri:2003ib}. On the other hand, in the $AdS_3\times S^3\times S^3 \times S^1$ case the dual gauge theory remains largely unknown, mainly due to the fact that the supergravity approximation fails to be as useful as in the other examples \cite{Gukov:2004ym}. Even though the dual gauge theory remains illusive it is still possible to investigate the integrable structures of the theory. This is achieved by formulating the string as a supercoset sigma model whose classical equations of motion allow for a Lax representation ensuring classical integrability \cite{Bena:2003wd,Babichenko:2009dk}. By integrating the Lax connection around a closed loop one gets the monodromy matrix which in turn can be used to generate an infinite tower of conserved charges. Furthermore, the finite gap method can be used to reformulate the equations of motion in terms of a set of integral equations \cite{Kazakov:2004qf,Beisert:2005bm}. These integral equations in turn arise as the semiclassical limit of a set of conjectured quantum Bethe equations \cite{Staudacher:2004tk,Beisert:2005tm}. However, the supercoset sigma models underlying this construction suffer from a serious drawback: They don't always describe all the physical (fermionic) degrees of freedom of the superstring. The reason for this is that in \3 or in $AdS_4\times\mathbbm{CP}^3$ the supercoset sigma model corresponds to a certain kappa-symmetry gauge-fixed version of the full Green-Schwarz superstring \cite{Arutyunov:2008if,Gomis:2008jt,Rughoonauth:2012qd}. This gauge-fixing breaks down for certain configurations of the string \cite{Grassi:2009yj,Cagnazzo:2009zh,Rughoonauth:2012qd}, in particular when the string moves only in the $AdS$ subspace, causing the supercoset sigma model to be missing physical degrees of freedom in these cases. The $AdS_2\times S^2\times T^6$ string is an even more striking example since in this case the supercoset model is merely a consistent truncation of the full string action and never describes all the degrees of freedom \cite{Sorokin:2011rr}. It is therefore of interest to try to generalize the techniques used for supercoset models to the full string action without kappa-symmetry fixing. A first step in this direction is to try to construct a Lax connection for the string without fixing kappa-symmetry. This has been done in \cite{Sorokin:2010wn} for the $AdS_4\times\mathbbm{CP}^3$ string and in \cite{Sorokin:2011rr} for the $AdS_2\times S^2\times T^6$ string to quadratic order in fermions (see also \cite{Cagnazzo:2011at} and, for a somewhat different approach \cite{Uvarov:2012bh}). Here we will show that the same procedure works also for the \3 string. The next step would be to formulate the Bethe ansatz equations. It is not yet known how to do this in the general (non-supercoset) case.

The Bethe equations are derived from properties of an underlying global symmetry which for \3 is $d(2,1;\alpha)^2$ (this corresponds to a subgroup of the superisometry group of the background $D(2,1;\alpha)^2\times U(1)$). Here the parameter $\alpha$ is defined through the triangle identity \cite{Babichenko:2009dk},
\bea 
\label{eq:triangel-identity}
\frac{1}{R^2}=\frac{1}{R_+^2}+\frac{1}{R_-^2},\qquad \alpha=\frac{R^2}{R_+^2}=\cos^2\phi\,,
\eea 
where $R$ is the AdS and $R_\pm$ the three-sphere radii. At the special points $\alpha=0,1$ or equivalently $\phi=0,\frac{\pi}{2}$ \3 'decompactifies' to $AdS_3\times S^3 \times T^4$ and thus one unified description parameterized by the angle $\phi$ covers both cases. In the $T^4$ case the global symmetries becomes $psu(1,1|2)^2=d(2,1;0)^2=d(2,1;1)^2$ \cite{Babichenko:2009dk}.

For $\phi\neq0,\frac{\pi}{2}$ the string sigma model has 6 massive and 2 massless bosonic modes plus 8, generically massive, fermionic modes while in the $T^4$ case the spectrum decomposes into $4+4$ massive and $4+4$ massless modes. The massless modes are a novel feature in $AdS / CFT$ and as mentioned above they are problematic to incorporate in the Bethe ansatz equations since they can not be addressed with the finite gap techniques. What is more, the Bethe equations are parameterized by $\phi$ and allow for a smooth limit between the two supergroups, but the underlying spin-chain is alternating for $d(2,1;\alpha)$ and homogeneous for $psu(1,1|2)$ \cite{OhlssonSax:2011ms}. Thus the 'decompactified' limit seems slightly ambiguous at the level of the spin-chain. 

There is also the notion of a composite heavy mode. For the alternating spin-chain the massive modes come with masses $1$, $\cos^2\phi$ and $\sin^2\phi$ in suitable units (each described by one complex field) and the heavy mode is conjectured to be a composite state made out of the two light ones, similar to $AdS_4 / CFT_3$. However, for $psu(1,1|2)$ this should not be the case since all excitations have the same mass, now similar to $AdS_5 / CFT_4$. Thus the composite nature should somehow disappear as $\phi\rightarrow0,\frac{\pi}{2}$. What happens at the level of the string Lagrangian? Here $\phi$ is also just a parameter and limits between the two backgrounds can be taken smoothly. Classically the heavy mode is just another string coordinate and if it is indeed composite, as suggested by the Bethe ansatz, then it should disappear as a fundamental excitation once quantum corrections are taken into account \cite{Zarembo:2009au,Sundin:2009zu,Zarembo:2011ag}. However, one should be able to find hints for this already at the level of the classical sigma model. For example, if the heavy mode is composite, one would expect a decay process of a heavy mode into two light ones. For the \3 string this is exactly what happens since the string Lagrangian has cubic interaction terms mediating this kind of process, once again similar to the $AdS_4\times \mathbbm{CP}^3$ case. What is more, for the special case of $T^4$ the cubic interaction terms completely disappear and the (classical) arguments for having a composite mode vanish. Thus, qualitatively the findings of the Bethe ansatz can be motivated from the sigma model. 

In this paper we will address in detail some of the issues mentioned above. Our starting point will be the type IIA \3 Green-Schwarz string Lagrangian derived in \cite{Rughoonauth:2012qd}. In section \ref{sec:classical-integrability} we prove that the (non-gauge-fixed) Green-Schwarz string is indeed classically integrable, at least up till quadratic order in fermions. Our construction covers at the same time also the $AdS_4\times\mathbbm{CP}^3$ and $AdS_2\times S^2\times T^6$ cases whose integrability has been discussed before. The proof involves writing the superisometry algebra in a suitable form and then constructing the Lax connection out of components of the worldsheet Noether currents of the superisometries. We also discuss the kappa-symmetry transformation properties of this Lax connection. In section \ref{sec:gaugefixed-BMN} we turn to a perturbative study starting from the quartic near BMN action of \cite{Rughoonauth:2012qd}. Integrability dictates a quantum dispersion relation determined up to an unknown interpolating function $h(\lambda)$ \cite{Babichenko:2009dk,OhlssonSax:2011ms}. In $AdS_4 / CFT_3$ this function turned out to depend on the regularization in the strong coupling regime $\lambda\gg1$. That is, depending on how the theory is regularized, different finite answers could be obtained. This can be traced back to the treatment of the heavy modes. Whether one uses a cutoff treating the massive modes as composite or not, a different one-loop correction to $h(\lambda)$ is obtained. If the heavy mode is taken as composite, an algebraic curve (AC) inspired cutoff is natural. On the other hand, from the worldsheet (WS) point of view where each string coordinate is treated on equal footing another cutoff is more natural. In section \ref{sec:quantumdispersion} we determine this subleading contribution for the \3 string, in both AC and WS regularizations, by computing the one-loop contribution to propagators of the heavy, light and massless bosonic string modes. Interestingly, for equal $S^3$ radii we find that the result is identical to the strong coupling results of $AdS_4 / CFT_3$. We end the paper with several appendices explaining some of the technical parts of the computations.

\section{Integrability of the Green-Schwarz string in certain $AdS$ backgrounds}
\label{sec:classical-integrability}
In this section we show that the Green-Schwarz superstring in $AdS_3\times S^3\times S^3\times S^1$ is classically integrable (without any kappa symmetry gauge-fixing), at least up to quadratic order in fermions, by constructing its Lax connection from components of the Noether currents of the superisometries of the background. The construction covers also the $AdS_4\times\mathbbm{CP}^3$ and $AdS_2\times S^2\times T^6$ case considered previously in \cite{Sorokin:2010wn} and \cite{Sorokin:2011rr} and the proof is essentially the same although we've been slightly more general in order to cover all cases at once. The only really new material here is the casting of the superisometry algebra of $AdS_3\times S^3\times S^3\times S^1$ in the appropriate form, which is described in Appendix A, and the discussion of the kappa-symmetry variation of the Lax connection given at the end of this section. Since much of the material is discussed in more detail in the papers mentioned above we will be rather brief here. Let us also mention that it should be straight-forward to construct the Lax connection to all orders in the coset fermions and quadratic order in the non-coset fermions, as was done for the $AdS_4\times\mathbbm{CP}^3$ and  $AdS_2\times S^2\times T^6$ string in \cite{Cagnazzo:2011at}, but we will not do this here.

\subsection{Green-Schwarz action to quadratic order in fermions}
We will consider only type IIA supergravity backgrounds here but everything we say extends in a simple way to type IIB. The action for the GS superstring in a type IIA supergravity background
(with zero background fermionic fields and NS--NS flux, and constant dilaton $\phi_0$ (not to be confused with the angle $\phi$)) takes the following form up to quadratic
order in fermions (the two Majorana-Weyl spinors of type IIA superspace are described as a single 32-component Majorana spinor $\Theta$) \cite{hep-th/9601109,hep-th/9907202}\footnote{In the context of AdS/CFT the string tension is proportional to the 't Hooft coupling $\sqrt{\lambda}$.}
\begin{equation}
\label{action}
S=-T\int\left(\frac{1}{2}*e^Ae_A+i*e^A\,\Theta\Gamma_A{\mathcal D}\Theta-ie^A\,\Theta\Gamma_A\Gamma_{11}{\mathcal D}\Theta\right)\,,
\end{equation}
where the $e^A(X)$ $(A=0,1,\cdots,9)$ are worldsheet pullbacks of the vielbein one-forms of the purely bosonic part of the background ($*$ denotes the worldsheet Hodge-dual and we leave the wedge product implicit), and the generalized covariant derivative acting on the fermions is given by
\begin{equation}
\label{E}
{\mathcal D}\Theta=(\nabla-\frac{1}{8}e^A\,\slashed F\Gamma_A)\ \Theta\quad\mbox{where}\quad \nabla\Theta=(d-\frac{1}{4}\omega^{AB}\Gamma_{AB})\Theta\,,
\end{equation}
where $\omega^{AB}$ is the spin connection of the background space-time and the coupling to the RR fields comes from the matrix
\begin{equation}
\label{eq:slashedF}
\slashed F=e^{\phi_0}\left(-\frac{1}{2}\Gamma^{AB}\Gamma_{11}F_{AB}+\frac{1}{4!}\Gamma^{ABCD}F_{ABCD}\right)\,.
\end{equation}

\subsection{Superisometry currents}
We are interested in backgrounds with superisometries. We therefore assume our string action to be invariant under the transformations
\begin{equation}
\delta X^M e_M{}^A(X)=K^A(X)+i\Theta\Gamma^A\Xi\,,\qquad\delta\Theta=\Xi-\frac{1}{4}\Gamma^{AB}\Theta\,(\nabla_AK_B-K^M\omega_{MAB})\,,
\end{equation}
where $K_A$ are Killing vectors, so that
\begin{equation}
\label{eq:Killing-vector}
\nabla_{(A}K_{B)}=0\qquad\Rightarrow\qquad 
\nabla_C\nabla_AK_B
%=2\nabla_(A\nabla_|B|K_C)-\nabla_C\nabla_BK_A
=R_{ABC}{}^DK_D\,,
\end{equation}
where $R_{AB}{}^{CD}$ is the Riemann tensor and $\Xi$ are Killing spinors satisfying the Killing spinor equation
\begin{equation}
\label{eq:Killingspinor}
{\mathcal D}\Xi=\nabla\Xi-\frac{1}{8}e^A\,\slashed F\Gamma_A\Xi=0\,.
\end{equation}

The Noether current corresponding to supersymmetry transformations (i.e. the $\Xi$-isometries) is easily seen to be given by
\begin{equation}
\label{eq:Jsusy}
J_{susy}=\frac{i}{2R}\left(e^A\,\Theta\Gamma_A\Xi-*e^A\,\Theta\Gamma_A\Gamma_{11}\Xi\right)\,,
\end{equation}
where the normalization is chosen for later convenience ($R$ will be the $AdS$-radius). The fact that this current is conserved, i.e.
\begin{equation}
d*J_{susy}=0
\end{equation}
is easily seen to follow from the Killing spinor equation and the equations of motion for $\Theta$ and $X$ which take the form
\begin{equation}
\label{eq:eom}
i*e^A\,\Gamma_A{\mathcal D}\Theta-ie^A\,\Gamma_A\Gamma_{11}{\mathcal D}\Theta=0\,,\qquad \nabla*e^A=\Theta^2\mbox{-terms}\,.
\end{equation}
Remember that we are dropping all term of higher that quadratic order in fermions.

The Noether current corresponding to the bosonic isometries takes the form
\begin{equation}
\label{eq:Jb}
J_{\mathcal B}=J^AK_A+J^{AB}\nabla_AK_B=e^A\,K_A+\mbox{fermions}\,,
\end{equation}
where the two pieces are given by
\begin{eqnarray}
J^A&=&e^A
+i\Theta\Gamma^A{\mathcal D}\Theta
+i\Theta\Gamma^A\Gamma_{11}*{\mathcal D}\Theta
-\frac{i}{8}e^B\,\Theta\Gamma_B\slashed F\Gamma^A\Theta
+\frac{i}{8}*e^B\,\Theta\Gamma_B\Gamma_{11}\slashed F\Gamma^A\Theta
\nonumber\\
J^{AB}&=&
-\frac{i}{4}e^C\,\Theta\Gamma^{AB}{}_C\Theta
+\frac{i}{4}*e^C\,\Theta\Gamma^{AB}{}_C\Gamma_{11}\Theta\,.
\label{eq:JA-JAB}
\end{eqnarray}
Its conservation, i.e. $d*J_{\mathcal B}=0$, again follows from the equations of motion for $X$ and $\Theta$ together with the fact that $K_A$ are Killing vectors. The conservation of $J_{\mathcal B}$ is equivalent to the equations for the components
\begin{equation}
\label{eq:cons-comp}
\nabla*J^A+*e^D\,J^{BC}R_{BCD}{}^A=0\,,\qquad\nabla*J^{AB}+*e^{[A}\,J^{B]}=0\,.
\end{equation}
These equations will be useful when we check the flatness of the Lax connection to be discussed below.

\subsection{Classical integrability of the GS string in certain backgrounds}
In this section we will show that in certain $AdS$-backgrounds, where $\slashed F$ and the superisometry algebra take a specific form, it is possible to construct a one-parameter family of flat connections, i.e. a Lax connection, for the Green-Schwarz string (still at quadratic order in fermions). Our first assumption will be that $\slashed F$, defined in (\ref{eq:slashedF}), should take the following form
\begin{equation}
\label{eq:FP}
\slashed F=-\frac{4i^k}{R}\mathcal P\Gamma_*\,,\qquad \Gamma_*=i^k\Gamma\cdots\Gamma\,,\qquad\Gamma_*^2=1\qquad(k=0\quad\mbox{or}\quad1)\,,
\end{equation}
where $R$ is the $AdS$-radius, $\Gamma_*$ is a product of gamma-matrices squaring to one (see (\ref{eq:table}) below) and $\mathcal P$ is a projection operator which singles out the supersymmetries of the background, i.e. the parameters in the supersymmetry transformations satisfy $\epsilon=\mathcal P\epsilon$ and similarly for the Killing spinors $\Xi$ defined in (\ref{eq:Killingspinor}). From the definition of $\slashed F$ in (\ref{eq:slashedF}) it follows that
\begin{equation}
[\mathcal P\Gamma_*,\Gamma_{11}]=0\qquad\mbox{and}\qquad(\mathcal C\mathcal P\Gamma_*)_{\alpha\beta}=-(\mathcal C\mathcal P\Gamma_*)_{\beta\alpha}\,,
\end{equation}
where we have indicated the charge conjugation matrix $\mathcal C$ explicitly for clarity. Defining
\begin{equation}
\bar{\mathcal P}=-\mathcal C\mathcal P^T\mathcal C\qquad\Rightarrow\qquad \mathcal P\Gamma_*=\mp\Gamma_*\bar{\mathcal P}\qquad\mbox{for}\qquad (\mathcal C\Gamma_*)_{\alpha\beta}=\pm(\mathcal C\Gamma_*)_{\beta\alpha}\,.
\end{equation}
Furthermore the integrability condition for the Killing spinor equation (\ref{eq:Killingspinor}) requires that
\begin{equation}
(1-\mathcal P)\Gamma^{AB}\mathcal P\,R_{AB}{}^{CD}=0\,.
\end{equation}

The second assumption we will make is that the algebra of superisometries can be brought to the following form
\begin{eqnarray}
[P_A,P_B]=-\frac{1}{2}R_{AB}{}^{CD}M_{CD}\,,\qquad[M_{AB},P_C]=\eta_{AC}P_B-\eta_{BC}P_A\,,
\nonumber\\
{}[M_{AB},M_{CD}]=\eta_{AC}M_{BD}+\eta_{BD}M_{AC}-\eta_{BC}M_{AD}-\eta_{AD}M_{BC}\,,
\label{eq:PMalgebra}
\end{eqnarray}
where $R_{AB}{}^{CD}$ is the Riemann tensor of the space in question and
\begin{eqnarray}
[P_A,Q]=\frac{i^k}{2R}Q\Gamma_*\Gamma_A\mathcal P\,,\qquad[M_{AB},Q]=-\frac{1}{2}Q\Gamma_{AB}\mathcal P
\nonumber\\
\{Q,Q\}=2i(\mathcal C\bar{\mathcal P}\Gamma^A\mathcal P)\,P_A+\frac{i^{1-k}R}{2}(\mathcal C\bar{\mathcal P}\Gamma^{AB}\Gamma_*\mathcal P)\,R_{AB}{}^{CD}M_{CD}\,,
\label{eq:Qalgebra}
\end{eqnarray}
where the supersymmetry generators have the appropriate projection, i.e. $Q=Q\mathcal P$. The relation between the generators $P_A$, $Q$ and the Killing vector and Killing spinor $K_A$, $\Xi$ will be explained below.

We will now list some examples of type IIA supergravity solutions for which these assumptions hold, i.e. the algebra of superisometries can be brought to the form (\ref{eq:PMalgebra}) and (\ref{eq:Qalgebra}) and $\slashed F$ takes the form in (\ref{eq:FP}). The spaces together with the fluxes, the form of $\Gamma_*$ (for an appropriate choice of coordinates) and the number of supersymmetries they preserve are
\begin{equation}
\begin{array}{l|ccc}
&\mbox{Fluxes}& \Gamma_* & \#\mbox{SUSYs}\\
\hline & & &\\
AdS_2\times S^2\times T^6 & F_2(AdS_2),\quad F_4(S^2\times T^6)&\Gamma^{01}\Gamma_{11} &8\\
 & & &\\
AdS_2\times S^2\times T^6 & F_2(S^2),\quad F_4(AdS_2\times T^6)&i\Gamma^{01}\Gamma^{456789}&8\\
 & & &\\
AdS_2\times S^2\times T^6 & F_4(AdS_2\times S^2\times T^4)& -\Gamma^{23}\Gamma^{46}&8\\
 & & &\\
AdS_3\times S^3\times S^3\times S^1 & F_4(AdS_3\times S^3\times S^3\times S^1)&i\Gamma^{012}\Gamma^9 &16\\
 & & &\\
AdS_4\times\mathbbm{CP}^3 & F_2(CP^3),\quad F_4(AdS_4) & i\Gamma^{0123} &24\\
 & & &\\
\end{array}
\label{eq:table}
\end{equation}
The three $AdS_2$ solutions are related by T-duality and the classical integrability of the GS string in these backgrounds was discussed in \cite{Sorokin:2011rr}. The integrability of the GS string in $AdS_4\times\mathbbm{CP}^3$ was discussed in \cite{Sorokin:2010wn}. For the $AdS_3\times S^3\times S^3\times S^1$ case the fact that $\slashed F$ takes the form (\ref{eq:FP}) was shown in \cite{Rughoonauth:2012qd}\footnote{Note that in this case $\bar{\mathcal P}=1-\mathcal P$.}. The fact that the superisometry algebra can be brought to the form (\ref{eq:PMalgebra}) and (\ref{eq:Qalgebra}) also in this case is demonstrated in Appendix A. It is also worth mentioning that the $AdS_3\times S^3\times S^3\times S^1$ case really describes a one-parameter family of solutions with the parameter $\phi$ relating the radii of the two $S^3$'s according to (\ref{eq:triangel-identity}). It includes the special case $AdS_3\times S^3\times T^4$ corresponding to $\phi=0\,(\pi/2)$.

\subsubsection*{Constructing the Killing vectors and Killing spinors}
We will assume that the bosonic subspace of the background is a symmetric space, as is clearly the case for the spaces listed above. This guarantees that the purely bosonic part of the string action is classically integrable since it can be formulated as a coset sigma model. This also allows us to construct the Killing vectors and Killing spinors from the generators of the superisometry algebra. Let $g$ be an element of bosonic subgroup of the superisometry group and define the Maurer-Cartan form as
\begin{equation}
\label{eq:MCform}
K=g^{-1}dg=\frac{1}{2}\omega^{AB}M_{AB}+e^AP_A\,,\qquad\Rightarrow\qquad dK=KK\,. %\,,\qquad g\in SO(2,2)\times SO(4)\times SO(4)\times U(1)
\end{equation}
This defines the bosonic vielbeins and spin connection of our background. The Killing vectors are then defined as
\begin{equation}
K_A=gP_Ag^{-1}\,,
\end{equation}
which gives
\begin{equation}
\nabla_AK_B=[K_A,K_B]\,.
%\nabla_AK_B=d_AK_B-\omega_{AB}{}^CK_C
\end{equation}
From the commutator of $P$ with itself in (\ref{eq:PMalgebra}) we also conclude that
\begin{equation}
-\frac{1}{2}R_{AB}{}^{CD}gM_{CD}g^{-1}=\nabla_AK_B\qquad\Rightarrow\qquad[K_C,\nabla_AK_B]=R_{ABC}{}^DK_D\,.
\end{equation}
The Killing spinors are defined as
\begin{equation}
\label{eq:defXi}
\Xi=i^{1+k}gQ\Gamma_*\mathcal Cg^{-1}\,.
\end{equation}
From (\ref{eq:Qalgebra}) it then follows that
\begin{equation}
\label{eq:XiXi}
\{\Xi,\Xi\}=-2i^{1+2k}(\mathcal P\Gamma_*\Gamma^A\Gamma_*\bar{\mathcal P}\mathcal C)\,K_A+i^{1+k}R(\mathcal P\Gamma_*\Gamma^{AB}\bar{\mathcal P}\mathcal C)\,\nabla_AK_B
\end{equation}
and
\begin{eqnarray}
\label{eq:KXi}
[\nabla_AK_B,\Xi]=-\frac{1}{4}R_{AB}{}^{CD}\,\Gamma_*\Gamma_{CD}\Gamma_*\Xi\,,\qquad
%[K_A,\Xi]=-\frac{i^k}{2R}\Xi\Gamma_A\mathcal P\gamma_*C
[K_A,\Xi]=-\frac{i^k}{2R}\mathcal P\Gamma_*\Gamma_A\Xi\,.
%\qquad\Rightarrow\qquad
%\nabla\Xi=e^A\,[K_A,\Xi]=-\frac{i^k}{2R}e^A\,\Xi\Gamma_A\mathcal P\Gamma_*C\,.
\end{eqnarray}
It follows from this equation and the definition (\ref{eq:defXi}) and (\ref{eq:MCform}) that
\begin{equation}
\nabla\Xi=-\frac{i^k}{2R}e^A\,\mathcal P\Gamma_*\Gamma_A\Xi\,,
\end{equation}
which, using (\ref{eq:FP}), is precisely the Killing spinor equation (\ref{eq:Killingspinor}). This completes the construction of the Killing vectors and Killing spinors in terms of the generators of the superisometry group.

\subsubsection*{The Lax connection}
The first step to showing the classical integrability will be to show that the following two highly non-trivial identities hold
\begin{eqnarray}
dJ_{susy}&=&-2(J_{\mathcal B}J_{susy}+J_{susy}J_{\mathcal B})
\nonumber\\
(\nabla J^{AB}+(J-e)^Ae^B)\nabla_AK_B&=&-J_{susy}^2\,.
\label{eq:intcond}
\end{eqnarray}
The left-hand-side of the first equation is easily computed using the form of the supersymmetry current in (\ref{eq:Jsusy}) together with the equations of motion for the fermions (\ref{eq:eom}) and the Killing spinor equation (\ref{eq:Killingspinor}), while the left-hand-side of the second follows from the form of the components of the bosonic supercurrent in (\ref{eq:JA-JAB}) and the equations of motion of the fermions
\begin{eqnarray}
dJ_{susy}&=&\frac{i}{8R}\left(e^Ae^B\,\Theta\Gamma_A\slashed F\Gamma_B\Xi-*e^Ae^B\,\Theta\Gamma_A\Gamma_{11}\slashed F\Gamma_B\Xi\right)
\nonumber\\
\left(\nabla J^{AB}+(J-e)^Ae^B\right)\nabla_AK_B&=&
\frac{i}{16}\left(
*e^Ce^D\,\Theta\Gamma_C\Gamma^{AB}\Gamma_{11}\slashed F\Gamma_D\Theta
-e^Ce^D\,\Theta\Gamma_C\Gamma^{AB}\slashed F\Gamma_D\Theta
\right)
\nabla_AK_B
\,.\nonumber
\end{eqnarray}
Using the form of $\slashed F$ in (\ref{eq:FP}), the form of $J_{susy}$ and the fact that $J_{\mathcal B}=e^AK_A+\mbox{fermions}$, as well as the algebra of the Killing spinors and Killing vectors in (\ref{eq:XiXi}) and (\ref{eq:KXi}), these expressions can be seen to agree precisely (up to quadratic order in fermions) with the right-hand-sides of (\ref{eq:intcond}). We will now see that these two relations guarantee the classical integrability.

We consider the following Lax connection, which is built out of components of the conserved superisometry currents defined in eqs. (\ref{eq:Jb}), (\ref{eq:JA-JAB}) and (\ref{eq:Jsusy}),
\begin{equation}
\label{eq:Lax}
L=\alpha_1e^A\,K_A+\alpha_2*J_{\mathcal B}+\alpha_2^2J^{AB}\,\nabla_AK_B+\alpha_1\alpha_2*J^{AB}\,\nabla_AK_B-\alpha_2\beta_1J_{susy}+\alpha_2\beta_2*J_{susy}\,.
\end{equation}
The four parameters appearing in $L$ satisfy the following three equations
\begin{equation}
\label{eq:spectral}
\alpha_2^2=2\alpha_1+\alpha_1^2\,,\qquad\beta_1=\mp\sqrt{\frac{\alpha_1}{2}}\,,\qquad\beta_2=\pm\frac{\alpha_2}{\sqrt{2\alpha_1}}\,,
\end{equation}
which means that they can all be expressed in terms of a single (spectral) parameter. Therefore eq. (\ref{eq:Lax}) defines a one-parameter family of one-forms, or connections, on the string worldsheet. What is remarkable is that these connections are actually flat (again up to quadratic order in fermions), i.e.
\begin{equation}
dL-L\wedge L=0\,.
\end{equation}
This is not very difficult to show and the calculation is essentially the same as in \cite{Sorokin:2010wn,Sorokin:2011rr}. Using the conservation of the superisometry currents, i.e. $d*J_{\mathcal B}=0=d*J_{susy}$ and (\ref{eq:cons-comp}), together with the important identities derived at the beginning of this section (\ref{eq:intcond}) and the Killing vector identities (\ref{eq:Killing-vector}) one finds that
\begin{eqnarray}
dL&=&
e^A\left(
(\alpha_2^2-\alpha_1)e^B
-\alpha_2^2J^B
+\alpha_1\alpha_2*J^B
\right)
\,\nabla_AK_B
-\alpha_2e^C\left(\alpha_2J^{AB}+\alpha_1*J^{AB}\right)\,\nabla_C\nabla_AK_B
\nonumber\\
&&{}
-\alpha_2^2J_{susy}^2
+2\alpha_2\beta_1(J_{\mathcal B}J_{susy}+J_{susy}J_{\mathcal B})\,.
\end{eqnarray}
On the other hand one finds, using the parameter relations in (\ref{eq:spectral}), dropping terms of higher than quadratic order in fermions and simplifying, that
\begin{eqnarray}
L\wedge L&=&
e^A\left(
(\alpha_2^2-\alpha_1)e^B
-\alpha_2^2J^B
+\alpha_1\alpha_2*J^B
\right)
\,[K_A,K_B]
\nonumber\\
&&{}
-\alpha_2e^C\left(\alpha_2J^{AB}+\alpha_1*J^{AB}\right)\,[K_C,\nabla_AK_B]
-\alpha_2^2J^2_{susy}
+2\alpha_2\beta_1(J_{\mathcal B}J_{susy}+J_{susy}J_{\mathcal B})\,.
\nonumber\\
\end{eqnarray}
Using the fact that $[K_A,K_B]=\nabla_AK_B$ and $\nabla_C\nabla_AK_B=R_{ABC}{}^DK_D=[K_C,\nabla_AK_B]$ we see that the Lax connection $L$ is indeed flat. The monodromy of $L$ can then be used to define a one-parameter family of conserved charges which demonstrates the integrability.

Let us also mention the fact that the form of the Lax connection in (\ref{eq:Lax}) somewhat obscures its invariance under the $\mathbbm Z_4$ automorphism of the superisometry algebra. As was shown in \cite{Cagnazzo:2011at} it is however possible to perform a gauge transformation of $L$ to bring it to a form which is manifestly $\mathbbm Z_4$ invariant.

We now turn to the question of how our Lax connection transforms under kappa-symmetry.

\subsubsection*{Kappa-symmetry variation of the Lax connection}
Since the Lax connection discussed in the previous section was constructed using components of the superisometry currents obtained from the GS string action, which we know is invariant under kappa-symmetry transformations, we would expect the Lax connection to transform nicely under kappa-symmetry. Here we will verify this explicitly (a proof based on general arguments was given in \cite{Cagnazzo:2011at}).

The GS string action (\ref{action}) is invariant under (local) kappa-symmetry transformations which (at linear order in $\Theta$) transform the target superspace coordinates as
\begin{eqnarray}
\label{eq:deltakappa}
\delta_\kappa X^Me_M{}^A=i\delta_\kappa\Theta\Gamma^A\Theta\,,\qquad \delta_\kappa\Theta=\frac{1}{2}(1+\Gamma)\kappa\,,
\qquad
\Gamma=\frac{1}{2\sqrt{-g}}\varepsilon^{ij}e_i{}^Ae_j{}^B\,\Gamma_{AB}\Gamma_{11}\qquad(\Gamma^2=1)\,,
\nonumber\\
\end{eqnarray}
where $g$ is the determinant of the induced metric $g_{ij}=e_i{}^Ae_j{}^B\eta_{AB}$ and $i,j=0,1$ are worldsheet indices. The action involves also an independent worldsheet metric $h_{ij}$, used to define the two-dimensional Hodge-dual, whose kappa-variation takes the form (see for example \cite{Grassi:2009yj})
\begin{equation}
\label{eq:deltahij}
\delta_\kappa(\sqrt{-h}h^{ij})=2i\sqrt{-h}\left(H^{ij}g^{kl}-2g^{k(i}h^{j)l}\right)
e_k{}^A\,\delta_\kappa\Theta\Gamma_A{\mathcal D}_l\Theta\,,
\end{equation}
where
\begin{eqnarray}
H^{ij}=h^{ij}-\frac{2h^{ik}g_{kl}h^{lj}-h^{ij}h^{kl}g_{kl}}{2\sqrt\frac{g}{h}+h^{kl}g_{kl}}
\quad\Rightarrow\quad h_{ij}H^{ij}=2\,,\qquad g_{ij}H^{ij}=2\sqrt\frac{g}{h}\,.
\end{eqnarray}
This somewhat complicated transformation of $h_{ij}$ ensures the invariance of the action. Since the Lax connection is only flat on-shell we are only interested in its kappa-variation on-shell. This means that we can set $h_{ij}=g_{ij}$ which means that $H^{ij}$ reduces to $h^{ij}$. Also, since our Lax connection is only valid to quadratic order in $\Theta$ we drop all terms beyond linear order in $\Theta$ in the kappa-variation.

Looking at the bosonic terms in the Lax connection we easily find
\begin{equation}
\delta_\kappa e^A=\nabla(i\delta_\kappa\Theta\Gamma^A\Theta)
\end{equation}
and
\begin{equation}
\delta_\kappa(e^A\,K_A)=
\nabla(i\delta_\kappa\Theta\Gamma^A\Theta)\,K_A
-ie^A\,\delta_\kappa\Theta\Gamma^B\Theta\,\nabla_AK_B\,.
%+\Theta^2
\end{equation}
We also need the kappa-variation of $*e^A$ which involves the worldsheet metric implicitly. Using its variation, recalling that on-shell $H^{ij}\rightarrow h^{ij}$ in (\ref{eq:deltahij}), it is possible to show that
\begin{equation}
\label{eq:deltaedual}
\delta_\kappa(*e^A)=
*\nabla(i\delta_\kappa\Theta\Gamma^A\Theta)
-2i\delta_\kappa\Theta\Gamma^A*\mathcal D\Theta
-2i\delta_\kappa\Theta\Gamma^A\Gamma_{11}\mathcal D\Theta\,,
\end{equation}
where the last two terms come from varying the metric implicit in the '$*$'. Unfortunately we have not found a simple proof of the fact that the last two terms take this form so the somewhat involved proof is deferred to Appendix B.

For the terms in the Lax connection (\ref{eq:Lax}) which involve fermions we only need to vary $\Theta$ since the other terms will be higher order. Using (\ref{eq:Jsusy}) we get
\begin{equation}
\delta_\kappa J_{susy}=
\frac{i}{2R}\left(
*e^A\,\Xi\Gamma_A\Gamma_{11}\delta_\kappa\Theta
-e^A\,\Xi\Gamma_A\delta_\kappa\Theta
\right)
%+\Theta^2-terms
=
-\frac{i}{2R}e^A\,\Xi\Gamma_A(1-\Gamma)\delta_\kappa\Theta=0\,.
\end{equation}
Here we have made use of the fact that $\delta_\kappa\Theta=\Gamma\delta_\kappa\Theta$ and that (on-shell)
\begin{equation}
\label{eq:kappaidentity}
e^A\Gamma_A\Gamma
%=1/2\sqrt{-h} \ve^{jk}e_i^Ae_j^Be_k^C\Gamma_A\Gamma_{BC}\Gamma_{11}
=
*e^A\,\Gamma_A\Gamma_{11}\,.
%+\Theta^2-terms
%*e^A=d\xi^i 1/\sqrt{-h} h_{ij}\ve^{jk}e_k^A
\end{equation}
Similarly we find varying (\ref{eq:JA-JAB}) and using the above identity that
\begin{eqnarray}
\delta_\kappa J^{AB}&=&
%-\frac{i}{2}e^C\,\Theta\Gamma^{AB}{}_C\delta\Theta
%+\frac{i}{2}*e^C\,\Theta\Gamma^{AB}{}_C\Gamma_{11}\delta\Theta
%+\Theta^2
%&=&
-\frac{i}{2}e^C\,\Theta\Gamma^{AB}{}_C(1-\Gamma)\delta_\kappa\Theta
-ie^{[A}\,\Theta\Gamma^{B]}\delta_\kappa\Theta
+i*e^{[A}\,\Theta\Gamma^{B]}\Gamma_{11}\delta_\kappa\Theta
\nonumber\\
&=&
-ie^{[A}\,\Theta\Gamma^{B]}\delta_\kappa\Theta
+i*e^{[A}\,\Theta\Gamma^{B]}\Gamma_{11}\delta_\kappa\Theta\,,
\end{eqnarray}
and
\begin{eqnarray}
\delta_\kappa(J-e)^A&=&
i\Theta\Gamma^A\nabla\delta_\kappa\Theta
+i\delta_\kappa\Theta\Gamma^A\nabla\Theta
+i\Theta\Gamma^A\Gamma_{11}*\nabla\delta_\kappa\Theta
+i\delta_\kappa\Theta\Gamma^A\Gamma_{11}*\nabla\Theta
\nonumber\\
&&{}
-\frac{i}{4}e^B\,\Theta\Gamma_B\slashed F\Gamma^A\delta_\kappa\Theta
+\frac{i}{4}*e^B\,\Theta\Gamma_B\Gamma_{11}\slashed F\Gamma^A\delta_\kappa\Theta\,.
%+\Theta^2
\end{eqnarray}
%where we have again made use of (\ref{eq:kappaidentity}).
Using these variations in the expression for the Lax connection (\ref{eq:Lax}) its variation (at linear order in $\Theta$) becomes
\begin{eqnarray}
\delta_\kappa L&=&
\alpha_1\delta_\kappa(e^A\,K_A)
+\alpha_2\delta_\kappa(*e^A\,K_A)
+\alpha_2*\delta_\kappa(J-e)^A\,K_A
+\alpha_2^2\delta_\kappa J^{AB}\,\nabla_AK_B
\nonumber\\
&&{}
+(1+\alpha_1)\alpha_2*\delta_\kappa J^{AB}\,\nabla_AK_B
%+\Theta^2
\nonumber\\
&=&
d\Lambda_\kappa+[L,\Lambda_\kappa]\,,
\end{eqnarray}
where
\begin{eqnarray}
\Lambda_\kappa=
-i\alpha_1\Theta\Gamma^A\delta_\kappa\Theta\,K_A
+i\alpha_2\Theta\Gamma^A\Gamma_{11}\delta_\kappa\Theta\,K_A\,.
\end{eqnarray}
We conclude that under kappa-symmetry the Lax connection transforms by a gauge-transformation, with parameter $\Lambda_\kappa$, which guarantees that it remains flat.

\section{Gauge fixed BMN Lagrangian}
\label{sec:gaugefixed-BMN}
Having established that the \3 string is classically integrable we now want to investigate its quantum properties in a BMN like expansion \cite{Berenstein:2002jq}. The starting point is to fix the worldsheet gauge invariance of (\ref{action}). We have already discussed the kappa-symmetry which can be used to gauge away half of the fermionic degrees of freedom. However, there are also bosonic symmetries in the form of Weyl and parameterization invariance on the worldsheet which can be used to eliminate some of the bosonic degrees of freedom. Fixing all the worldsheet gauges results in $8+8$ transverse bosonic and fermionic degrees of freedom. The \3 gauge fixed action was derived up till quartic order in fields (but only quadratic in fermions) in \cite{Rughoonauth:2012qd} and we will quickly review some key aspects of it in this section.

Expanding the action (\ref{action}) in transverse fields gives the BMN expansion \cite{Frolov:2006cc,Hentschel:2007xn,Astolfi:2008ji,Sundin:2008vt}
\bea \nn 
\mathcal{L}=\mathcal{L}_2+g^{-1/2}\mathcal{L}_3+g^{-1}\mathcal{L}_4+...
\eea 
where $g\sim \sqrt{\lambda}$ and is left implicit in most of the calculations. The quadratic piece can be solved exactly and is given by ($\partial_\pm=\partial_0\pm \partial_1)$
\bea 
\label{bmncoordinateL}
\mathcal{L}_2= i\bar\chi_+^i\partial_- \chi_+^i+i\bar\chi_-^i\partial_+\chi_-^i
+\frac{1}{2}\partial_+ y_i \partial_-\bar{y}_i
+\frac{1}{2}\partial_- y_i \partial_+\bar{y}_i
-m_i^2 y_i\bar{y}_i
-m_i\big(\bar\chi_+^i \chi_-^i+\bar\chi_-^i\chi_+^i\big)
\eea 
where 
\begin{equation}
\label{eq:spectrum}
m_1=1,\qquad m_2=\cos^2\phi,\qquad 
m_3=\sin^2\phi,\qquad m_4=0
\end{equation}
and $\phi$ parameterizes the relative size of the $S^3$ radii according to (\ref{eq:triangel-identity}). Thus we have four (complex) coordinates with generally distinct masses. One important simplifying limit is when $\phi=\frac{\pi}{4}$, corresponding to equal $S^3$ radii. In this limit the masses of the light coordinates $m_2=m_3=\frac{1}{2}$ and the spectrum resembles that of the $AdS_4\times \mathbbm{CP}^3$ string. This special limit will turn out to be quite useful in the following since it simplifies otherwise rather complicated expressions (without reducing to the even simpler $AdS_3\times S^3\times T^4$ case). 

The cubic piece of the Lagrangian is more complicated and is given by \cite{Rughoonauth:2012qd}
\begin{align}
\label{L3-full1}
\nonumber
\mathcal{L}_3 &= \frac{1}{2\sqrt{2}}\sin 2\phi\Big[-\cos^2\phi\big(\bar\chi_-^4\bar\chi_-^2-\bar\chi_-^1\bar\chi_-^3
+\bar\chi_+^1\bar\chi_+^3-\bar\chi_+^4\bar\chi_+^2\big)y_2 \\
& \phantom{\frac{1}{2\sqrt{2}}\sin 2\phi\Big[\quad} -i\sin^2\phi\big(\chi_-^3\bar\chi_-^4+\chi_-^2\bar\chi_-^1+\chi_+^3\bar\chi_+^4+\chi_+^2\bar\chi_+^1\big)y_3\\ \nn
& \phantom{\frac{1}{2\sqrt{2}}\sin 2\phi\Big[\quad} -2\big(\chi_-^2\bar\chi_+^3+\chi_+^2\bar\chi_-^3\big)y'_1 + 2\big(\chi_-^2\bar\chi_+^2-\chi_+^3\bar\chi_-^3\big)\dot{y}_4 \\ \nn
& \phantom{\frac{1}{2\sqrt{2}}\sin 2\phi\Big[\quad} + \big(\chi_-^3\bar\chi_+^4-\chi_-^2\bar\chi_+^1\big)(\dot{y}_3+y'_3) + \big(\chi_+^3\bar\chi_-^4-\chi_+^2\bar\chi_-^1\big)(\dot{y}_3-y'_3)\\ \nn 
& \phantom{\frac{1}{2\sqrt{2}}\sin 2\phi\Big[\quad} + i\big(\bar\chi_-^3\bar\chi_+^1+\bar\chi_-^2\bar\chi_+^4\big)(\dot{y}_2+y'_2) + i\big(\bar\chi_-^1\bar\chi_+^3+\bar\chi_-^4\bar\chi_+^2\big)(\dot{y}_2-y'_2)\Big]\\ \nn
& \phantom{\quad} -\frac{1}{\sqrt{2}}\sin 2\phi\ \big(\cos^2\phi\ |y_2|^2-\sin^2\phi\ |y_3|^2\big)\ \dot{y}_4 + \text{h.c.}\ ,
\end{align}
where the hermitian conjugate is defined in the standard way, $\big(\chi_-\bar\chi_+\big)^\dagger = \chi_+\bar\chi_-$. Here time $\partial_0$ and spatial $\partial_1$ derivatives are denoted by dots and primes respectively. Note that for the $\phi=0$ and $\phi=\pi/2$ cases, corresponding to $AdS_3\times S^3\times T^4$, the entire cubic Lagrangian collapses to zero.

Let us comment on what kind of three-vertex interactions we have between the light and the heavy modes. The decay processes possible for the heavy modes $y_1$ and $\chi_\pm^{(1)}$ are  ($\phi\neq 0,\pi/2$)
\begin{align} \label{eq:decayprocess}
&\textrm{Boson:}\qquad \qquad \parbox[top][0.8in][c]{1in}{\fmfreuse{decay1}} & \\ \nonumber &\textrm{Fermion:}\qquad \qquad \parbox[top][0.8in][c]{1.8in}{\fmfreuse{decay2}}
 \qquad \parbox[top][0.8in][c]{1in}{\fmfreuse{decay3}}
\end{align}
so the heavy mode can decay into two light ones, a property observed also for the heavy mode of the $AdS_4\times \mathbbm{CP}^3$ string \cite{Zarembo:2009au}. As we mentioned in the introduction, the Bethe ansatz treats the heavy mode as composite and the decay processes above seem to support, or at least not obviously contradict, this claim. However, since the entire cubic Lagrangian vanishes in the $\phi\rightarrow0\,(\pi/2)$, 'decompactifying', limit, so does the decay process. This again agrees with the Bethe ansatz solution (based on properties of $PSU(1,1|2)^2$) where the massive modes all have the same mass and all appear in the Bethe ansatz equations.

For completeness we also present the bosonic part of the quartic Lagrangian \cite{Rughoonauth:2012qd},
\begin{align}
\label{eq:l4_bosonig}
\nn
\mathcal{L}_4^{B} &= \frac{1}{4}\sin^2 2\phi\left(\cos^2\phi\ |y_2|^2-\sin^2\phi\ |y_3|^2\right)^2 - \frac{1}{8}\sin^2 2\phi\left(\dot{y}_4^2 + \dot{\bar{y}}_4^2 - y_4'^2 - \bar{y}_4'^2\right)\left(|y_2|^2 + |y_3|^2\right)\\ \nn 
&\phantom{\quad} - |\dot{y}_4|^2\left(|y_1|^2 - \cos 2\phi(\cos^2\phi|y_2|^2 - \sin^2\phi|y_3|^2)\right) + |\dot{y}_1|^2\big(\cos^4\phi|y_2|^2 + \sin^4\phi|y_3|^2\big) \\ \nn
&\phantom{\quad} - (|\dot{y}_2|^2 + |\dot{y}_3|^2 + |y'_i|^2)\left(|y_1|^2 - \cos^4\phi|y_2|^2 - \sin^4\phi|y_3|^2\right) - \cos^2\phi|\dot{y}_2|^2|y_2|^2-\sin^2\phi|\dot{y}_3|^2|y_3|^2\\ \nn
&\phantom{\quad} + \cos^2\phi\sin^2\phi|y'_4|^2(|y_2|^2 + |y_3|^2) - |y'_1|^2|y_1|^2 + \cos^2\phi|y'_2|^2|y_2|^2 + \sin^2\phi|y'_3|^2|y_3|^2\,.\\
\end{align} 
For the piece relevant for the one-loop computation in the next section, see Appendix C.

\section{Quantum dispersion relation}
\label{sec:quantumdispersion}
In this section we will calculate the leading quantum corrections to the dispersion relation of the bosonic string coordinates by evaluating the one-loop correction to two-point functions built out of the string modes using standard QFT perturbative techniques, see \cite{Abbott:2011xp,Giombi:2010bj} for similar analysis. To leading order the pole of the propagator is the standard relativistic one,
\bea \nn
p_0^2=E_i^2=p_1^2+m_i^2
\eea 
where $i=1,2,3,4$ is the coordinate label and the masses are given in (\ref{eq:spectrum}). Loop corrections enter as
\bea \label{eq:1looppole}
E_i^2=p_1^2+m_i^2+i\mathcal{A}^i+\ldots
\eea 
where $\mathcal{A}^i$ is the sum of one-loop diagrams contributing to the amplitude. Making the dependence on the coupling, $g=\sqrt\frac{\lambda}{2}$, explicit we have
\bea \label{eq:Epole}
E_i^2=M_i(\lambda)^2+\big[1+2\sqrt\frac{2}{\lambda}\,c^i\big]p_1^2+...,\qquad M_i(\lambda)^2=m_i^2+\sqrt\frac{2}{\lambda}\,q^i\,,
\eea  
where $c^i$ and $q^i$ are determined by explicitly calculating $\mathcal{A}^i$. $M_i(\lambda)$ is the renormalized mass defined by the $p_1\rightarrow 0$ limit. 

On the other hand, integrability implies an asymptotic dispersion relation of the form \cite{Babichenko:2009dk, OhlssonSax:2011ms, Beisert:2010jr}
\bea \label{eq:dispersionrelation}
E_k=\sqrt{m_k^2+4 h^2(\lambda) \sin^2\frac{p_1}{2}}
\eea 
where 
\bea \label{eq:interpolating}
h(\lambda)=\sqrt{\frac{\lambda}{2}}+c+\mathcal{O}(1/\sqrt{\lambda}),\qquad \lambda>>1
\eea 
is an interpolating function not determined by integrability, \cite{Nishioka:2008gz,Gaiotto:2008cg,Grignani:2008is,Gromov:2008fy}. The index $k$ here selects one of the two light coordinates, $k=2,3$. Assuming that the momentum scales as $p_1 \sim p_1/h(\lambda)$ and expanding in inverse powers of $h(\lambda)$ one finds that the first non-trivial correction is at two loops. If, on the other hand, one chooses a more standard expansion (at least from the string sigma model perspective) in inverse powers of the 't Hooft coupling, together with a similar scaling of the momentum, one finds
\bea 
E_k=\sqrt{m_k^2+p_1^2}+\sqrt{\frac{2}{\lambda}}\frac{cp_1^2}{\sqrt{m_k^2+p_1^2}}+\ldots
=\sqrt{m_k^2+\big[1+2\sqrt{\frac{2}{\lambda}}\,c\big]p_1^2}+\mathcal{O}(1/\lambda)\,.
\eea 
Thus, by computing (\ref{eq:1looppole}) we can recombine it into a square root dispersion relation and directly compare with the expression dictated by integrability. 

There are however a few complications along the way: Foremost, the BMN Lagrangian has cubic and quartic vertices, both giving rise to divergent loop integrals. Naturally, in order to obtain a well defined result a regularization scheme is needed, for example dimensional regularization or a momentum cutoff. How to incorporate the regularization is a bit subtle however. As it turns out, and as was also the case in $AdS_4/CFT_3$, the final result is regularization dependent. Regularization dependence means that we can get different finite terms in (\ref{eq:1looppole}) due to $\infty -\infty$ ambiguities. This can be traced back to how we treat the heavy modes of the theory. In the Bethe ansatz the heavy modes are composite states of two lighter ones, similar to $AdS_4 / CFT_3$. To avoid the possibility of a heavy mode decaying into nothing motivates using a mode number cutoff such that the decay process depicted in (\ref{eq:decayprocess}) is allowed up to the cutoff scale \cite{Astolfi:2011ju, Astolfi:2011bg}. On the other hand, from the string worldsheet point-of-view there is no distinction between the light and heavy modes and it's natural to use the same cutoff for both. The former we will call algebraic curve (AC) and the latter worldsheet (WS) regularization, see \cite{Gromov:2008fy,Astolfi:2011ju,McLoughlin:2008ms,Alday:2008ut,Krishnan:2008zs,McLoughlin:2008he,Bandres:2009kw,Shenderovich:2008bs,Abbott:2010yb,Beccaria:2012qd} for extensive references. Since it's not completely clear even in $AdS_4/CFT_3$ which regulator is physical we will present our final answers in both the AC and WS scheme, see however \cite{LopezArcos:2012gb} for some recent results. Let's first have a look at the AC regularization. If we use a hard cutoff, we want to choose it so that the decay process (\ref{eq:decayprocess}) is allowed (on-shell) at the cutoff scale \cite{Astolfi:2011bg}\footnote{The relativistic frequencies are given by $\omega_1(m)=\sqrt{1+m^2}, \,\omega_2(m)=\sqrt{\cos^4\phi+m^2}$ and $\omega_3(m)=\sqrt{\sin^4\phi+m^2}$.}
\bea \label{H_LL_energysol}
\omega_1(\Lambda_1)=\omega_2(\Lambda_2)+\omega_3(\Lambda_3)\,,\qquad \Lambda_1=\Lambda_2+\Lambda_3\,,
\eea 
where $\Lambda_i$ denotes the cutoff for the corresponding mode. The above is solved by taking
\bea \label{eq:AC-cutoff}
\textrm{AC}:\qquad \Lambda_1=2\Lambda,\qquad \Lambda_2=2\cos^2\phi \Lambda,\qquad \Lambda_3=2\sin^2\phi \Lambda,\qquad \Lambda_4=\alpha_4\Lambda\,,
\eea 
where we've introduced the factor of $2$ to have the canonical normalization of the AC cutoff of $AdS_4 / CFT_3$ when $\phi=\frac{\pi}{4}$. We leave the cutoff for the massless mode unspecified and as it turns out $\alpha_4$ always comes multiplied with $m_4$ and hence vanishes. For the WS regularization on the other hand we simply employ the same cutoff for all fields, 
\bea
\label{eq:WS-cutoff}
\textrm{WS}:\qquad \Lambda_i=\Lambda,\qquad i=1,2,3,4\,.
\eea 
At the special point $\phi=\frac{\pi}{4}$ we can incorporate both cutoffs by introducing a parameter $\beta$ such that
\bea 
\label{eq:gen-cutoff}
\phi=\frac{\pi}{4}:\qquad 
\Lambda_{heavy}=\beta\,\Lambda_{light}
\eea 
where $\beta=2$ corresponds to AC and $\beta=1$ corresponds to the WS cutoff. What is more, at the special points where the background 'decompactifies' into $AdS_3\times S^3 \times T^4$ the two cutoffs become equivalent since the additional massless modes do not propagate in any loops, see later sections for explicit expressions. Also, we should mention that the word cutoff is somewhat misleading. In the following we will not use a hard cutoff in our loop integrals, only a generalized version of dimensional regularization that incorporates the finite effects of the modified hard cutoff. 

\subsection{Regularization}
Since the regularization is important let us explain in some detail what we will do. We will use the following notation for integrals
\bea 
\label{eq:integrals}
I_{n}^{s}\big[\Delta\big] =\int d^{2}\ell\,\frac{(\ell^2)^{s}}{\left[\ell^{2}-\Delta\right]^{n}}
\eea
which can be evaluated either in dimensional regularization (DREG) or momentum / hard cutoff. The guiding principle should be to pick the regulator that preserves the most symmetries. The hard cutoff clearly breaks translation invariance while DREG is a more symmetric regulator\footnote{While the sigma model is not Lorentz invariant beyond the quadratic approximation, some terms are nevertheless individually invariant - a symmetry which is broken by introducing a momentum cutoff.}. However, it is well known that DREG breaks supersymmetry since the degrees of freedom for gauge and gaugino fields fail to match. For the sigma model at hand however we have no worldsheet gauge fields or spinors, only worldsheet even and odd scalars, and  thus supersymmetry should be manifest in DREG with $d=2-\epsilon$. However, a priori it is not clear how we can implement the AC regulator in dimensional regularization. For a hard cutoff it is easy to implement this: For the three-vertex topologies we use the same cutoff $\Lambda$ regardless of which two fields propagate in the loop \cite{Abbott:2011xp} while for tadpole topologies we use (\ref{eq:AC-cutoff}). If we expand the relevant integrals in some arbitrary cutoff, $\beta \Lambda$, and DREG we have
\bea \nn
&& I^0_1\big[\Delta\big]_{\beta\Lambda}=i\pi\big(\log\Delta-2\log\left(\beta\Lambda\right)\big),\qquad 
I^1_1\big[\Delta\big]_{\beta\Lambda}=i\pi\big(\beta^2\Lambda^2+\Delta\log\Delta-2\Delta\log\left(\beta\Lambda\right)\big), \\ \nn 
&& I^0_1\big[\Delta\big]_\epsilon=i\pi\big(\gamma-\frac{2}{\epsilon}+\log\pi+\log\Delta\big), \qquad 
I^1_1\big[\Delta\big]_\epsilon=i\pi\Delta\big(\gamma-\frac{2}{\epsilon}+\log \pi+ \log\Delta\big)
\eea 
The quadratic divergence is absent in DREG since it is proportional to a sum of surface integrals which vanish in non-integer dimensions \cite{Smirnov:2006ry}. From the above it becomes clear how we can generalize the DREG tadpole integrals to incorporate the generalized cutoff for heavy modes
\bea  \label{eq:genDREG} 
\textrm{Generalized $\epsilon$-reg}:  \, I^0_1\big[m_i^2\big]\rightarrow  I^0_1\big[m_i^2\big]-2i\pi \log \alpha_i,\quad
I^1_1\big[m_i^2\big]\rightarrow  I^1_1\big[m_i^2\big]-2i\pi m_i^2\log \alpha_i,
\eea 
where 
\bea \nn 
\alpha_1=2,\qquad \alpha_2=2\cos^2\phi,\qquad \alpha_3=2\sin^2\phi
\eea
which up to the unphysical $\Lambda^2$ term exactly reproduces the $\Lambda$-integrals in AC regularization\footnote{We can leave $\alpha_4$ unspecified since it never enters any explicit calculations in the generalized $\epsilon$-reg.}. Thus, in the following it will be understood that every time we evaluate a tadpole integral, we use the generalized integrals above for the AC regularization.

\subsection{One-loop corrected propagators}
We now turn to the actual computation of the one-loop corrections. There are three distinct diagrams we need to take into account: Three-vertex bubble and tadpole diagrams and four-vertex tadpoles. In all cases the three-vertex tadpoles will be zero (sometimes trivially and sometimes through delicate cancellations between fermionic and bosonic loop diagrams). For the heavy and the massless modes we will derive expressions for arbitrary $\phi$ while the expressions for the two light modes are rather complicated and for these we restrict to $\phi=\pi/4$ (and the much simpler $AdS_3\times S^3\times T^4$ case, $\phi=0,\frac{\pi}{2}$). 

Before presenting the detailed analysis we will collect our findings here. Using the notation of (\ref{eq:Epole}) we find in the AC regularization
\bea \nn
&& \textrm{\underline{AC-regularization}:} \\ \nn 
&&
M_1(\lambda)^2=1,\qquad c_1=\frac{1}{4\pi}\sin^2 2\phi\log\sin 2\phi, \\ \nn 
&& 
M_2(\lambda)^2=M_3(\lambda)^2=\frac{1}{4}+\sqrt\frac{2}{\lambda}\,\frac{\log 2}{16\pi},\qquad c_2=c_3=0\qquad \big(\textrm{evaluated at }\phi=\frac{\pi}{4}\big), \\ \nn 
&& 
M_4(\lambda)^2=0,\qquad c_4=\frac{-1+2\log \sin 2\phi}{8\pi}\sin^22\phi
\eea 
while in WS regularization we get
\bea \nn
&& \textrm{\underline{WS-regularization}:} \\ \nn 
&&
M_1(\lambda)^2=1,\qquad c_1=\frac{1}{\pi}\Big(\cos^2\phi \log \cos\phi+\sin^2\phi\log\sin\phi\Big), \\ \nn 
&& 
M_2(\lambda)^2=M_3(\lambda)^2=\frac{1}{4},\qquad c_2=c_3=-\frac{\log 2}{2\pi}\qquad \big(\textrm{evaluated at }\phi=\frac{\pi}{4}\big), \\ \nn 
&& 
M_4(\lambda)^2=0,\qquad c_4=\frac{-\frac{\sin^2 2\phi}{8}+\cos^2\phi \log \cos\phi+\sin^2\phi\log\sin\phi}{\pi}
\eea 
Intriguingly we have for $\phi=\frac{\pi}{4}$, the same one-loop correction to $h(\lambda)$, given by $c_{2,3}$, as in $AdS_4\times\mathbbm{CP}^3$. Furthermore, for the light modes the mass gets renormalized in the AC scheme which is rather strange. One would expect no mass renormalization since the Bethe ansatz solution seems to indicate that the masses are $\cos^2\phi$ or $\sin^2\phi$ without corrections\footnote{Perhaps one should consider coupling renormalization which would effectively look like a shift in $\phi$. We thank C. Sieg for an interesting discussion regarding this.}. Perhaps this can be seen as an argument for employing the same cutoff for all fields which would agree with \cite{LopezArcos:2012gb}. It would be interesting to check whether the one-loop corrections to the two light fermionic modes are the same as for the bosonic ones for $\beta=2$ regularization.

As a final remark let us note that in the 'decompactifying' $T^4$ limit, both the AC and WS regulators give no one-loop corrections. This is expected since both for the $AdS_5 \times S^5$ and $AdS_2\times S^2\times T^6$ string the one-loop corrections are identically zero \cite{Callan:2004ev,preparation}. 

\subsubsection{Heavy mode}
In this and the following sections we will describe in detail how we arrive at the one-loop corrections in the AC and WS regularizations.

The relevant diagrams for the heavy mode are bubbles and four-vertex tadpole diagrams. In principle we could also have three-vertex tadpoles, but these are trivially zero since (\ref{L3-full1}) is at most linear in $y_1$. Let's start the analysis by investigating the bubble diagrams which always have two light fermions propagating in the loop. Evaluating the diagram close to $p_0=\sqrt{1+p_1^2}$ (i.e. on-shell) gives
\begin{align}
\label{bubble-y1}
\mathcal{A}^1_{B} & =\qquad\parbox[top][0.8in][c]{1.5in}{\fmfreuse{bubble-y1}}= \\ \nn 
&
-\frac{1}{(2\pi)^2}\int_0^1 dx\Big[\big((-1+x)x+\frac{1}{4}\sin^22\phi\big)I^0_2\big[(\cos^2\phi-x)^2\big]
%&
%-\frac{\sin^2 2\phi\,p_1^2}{2\pi}\log\Lambda-\int^1_0 dx\,\frac{4x-2+\big(1-2x+\cos 2\phi\big)\log \frac{4}{(1-2x+\cos 2\phi)^2}}{4\pi(1-2x+\cos 2\phi)}\, p_1^2\sin^2 2\phi \\ \nn 
+I^1_2\big[(\cos^2\phi-x)^2\big]\Big] p_1^2\sin^2 2\phi\,.
%=i\frac{\big(\gamma+\log\pi-\frac{2}{\epsilon}\big)\sin^2 2\phi\,p_1^2}{4\pi}+i\frac{\log(\sin\phi\cos\phi)\,p_1^2\,\sin^22\phi}{2\pi}
\end{align}
The tadpole diagrams have fields of all flavors in the loop, 
\begin{align}
\label{tadpole-y1}
\mathcal{A}^1_{T} & =\qquad \parbox[top][0.8in][c]{1in}{\fmfreuse{tadpole-lightlight}}
= \\ \nn 
&\frac{1}{(2\pi)^2}\Big(\big(1+2p_1^2\big)I^0_1\big[1\big]-I^1_1\big[1\big]-2\cos^4\phi \,p_1^2 I^0_1\big[\cos^4\phi\big]-2\sin^4\phi \,p_1^2I^0_1\big[\sin^4\phi\big]\big)\Big)\,.
%\\ \nn 
%&=-i\frac{\big(\gamma+\log\pi-\frac{2}{\epsilon}\big)\sin^2 2\phi\,p_1^2}{4\pi}+\frac{i}{\pi}\Big(\log\beta+2\big(\cos^4\phi\log\cos\phi+\sin^4\phi\log\sin\phi\big)\Big)p_1^2
%+
%\parbox[top][0.8in][c]{1.5in}{\fmfreuse{tadpole-heavy}}
\end{align}
Both bubble and tadpole diagrams are separately divergent but their sum is manifestly finite but regularization dependent,
\bea \label{eq:y1oneloop}
\mathcal{A}^1_{AC}=-\frac{i}{2\pi}\sin^2 2\phi\,\log \big(\sin 2\phi \big)p_1^2,\quad 
\mathcal{A}^1_{WS}=-\frac{2i}{\pi}\Big(\cos^2\phi \log \cos\phi+\sin^2\phi\log\sin\phi\Big)p_1^2
\eea 
In the limiting case of equal $S^3$ radii at $\phi=\pi/4$ we find
\bea \label{eq:y1oneloopradii}
\mathcal{A}^1\vert_{\phi=\frac{\pi}{4}}= -\frac{i}{\pi}\log\frac{\beta}{2}\,p_1^2\,,
\eea
where $\beta=2\,(1)$ corresponding to AC(WS) regularization. This term is actually identical to the one-loop $\log$ correction to the heavy mode of the $AdS_4\times \mathbbm{CP}^3$ string (with $\beta=1$) \cite{Abbott:2011xp}. This is perhaps not very surprising since that model had similar three-vertex interactions as those in (\ref{eq:decayprocess}).

Another interesting limit is the 'decompactifying' limit $\phi=0$ (or equivalently $\phi=\pi/2$). In this case the one-loop correction is zero for both AC and WS schemes,
\bea 
\mathcal{A}^1\vert_{\phi=0,\frac{\pi}{2}}=0\,.
\eea 
This result is expected and agrees with the one-loop correction to the $AdS_5 \times S^5$ and $AdS_2\times S^2\times T^6$ strings, see \cite{preparation, INSPIRE-1112800}.

\subsubsection*{Composite nature of heavy mode?}
The Bethe ansatz treats the heavy mode as composite of the two light ones, similarly to what happens in $AdS_4 / CFT_3$. At the classical level of the sigma model, the only hint of this is the presence of cubic interaction terms mediating the decay processes depicted in (\ref{eq:decayprocess}). However, once quantum corrections are taken into account one should study the analytic properties of the two-point function for the heavy mode. In general the pole in the two-point function will get shifted by quantum corrections but remain, if this is the case we conclude that the state can be realized asymptotically and should be treated on equal footing with the other massive states. However it may also happen that the pole is replaced by a branch cut. In that case the corresponding particle does not exist as an asymptotic state. In \cite{Zarembo:2009au} it was suggested that this is precisely what happens for the heavy mode in $AdS_4\times\mathbbm{CP}^3$.

Let us quickly review the argument of \cite{Zarembo:2009au}. It turns out that the one-loop two-point function can be expanded close to the pole $\bar p^2 \sim 1$ as (assuming $\bar p^2<1$)
\bea \label{eq:offshellprop}
-i\mathcal{A}^1=a_0+a_{1/2}\sqrt{1-\bar p^2}+\mathcal{O}(1-\bar p^2),\qquad a_i=a_i(p_0,p_1)\,,
\eea 
where the coefficients are regular at $\bar p^2=1$. Now, if it happens that $a_0=0$, we find from (\ref{eq:1looppole}) that the equation for the one-loop corrected pole becomes
\begin{equation}
E^2-e(p_1)^2+a_{1/2}\sqrt{1-\bar p^2}+\ldots=0
\end{equation}
where $e(p_1)=\sqrt{p_1^2+1}$ is the classical pole. Or, since $E=e(p_1)+\mathcal O(\frac{1}{\sqrt\lambda})$,
\begin{equation}
e(p_1)-E-\frac{a_{1/2}}{\sqrt{2e(p_1)}}\sqrt{e(p_1)-E}\approx0\,.
\end{equation}
If $a_{1/2}<0$ this equation has no real solutions and we would conclude that the pole is replaced by a branch cut and the corresponding particle disappears from the asymptotic spectrum.

Now, let us see what happens in the case of $AdS_3\times S^3\times S^3\times S^1$. First we need the coefficients $a_0$ and $a_{1/2}$ in the expansion (\ref{eq:offshellprop}). The first one is easily read off from (\ref{eq:y1oneloop}), which shows that $a_0\neq0$ except for when $\phi=\frac{\pi}{4}$ and we use the AC regulator. To find the $a_{1/2}$ coefficient we need to evaluate the amplitude off-shell. It turns out that only the bubble diagram contributes and using the off-shell expression
\bea \nn 
&&
\mathcal{A}^1_B=-\frac{1}{(2\pi)^2}\int_0^1 dx\Big[	\big((-1+x)x\, \bar p^2+\frac{1}{4}\sin^2 2\phi\big) I^0_2\big[(1-x)\cos^4\phi+x\big(\bar p^2(-1+x)+\sin^4\phi\big)\big] \\ \nn 
&&
+I^1_2\big[(1-x)\cos^4\phi+x\big(\bar p^2(-1+x)+\sin^4\phi\big)\big]\Big]\sin^22\phi \,p_1^2
\eea 
one can show that 
\begin{equation}
a_{1/2}=-\frac{\sin2\phi\,p_1^2}{8}
\end{equation}
which is indeed negative in the whole interval $0<\phi<\frac{\pi}{2}$. When $\phi=\frac{\pi}{4}$ and we use the AC regulator, so that $a_0=0$, we can apply the above argument and conclude that it indeed appears like the heavy particle disappears from the spectrum due to quantum corrections. For generic values of $\phi$ we have $a_0\neq0$ but $a_{1/2}$ still has the correct sign. This happens also for spinning strings in $AdS_5\times S^5$ \cite{Giombi:2010bj} and seemed to indicate a discrepancy with the Bethe ansatz solution, however a more sophisticated argument for the disappearance of the heavy pole was given in \cite{Zarembo:2011ag} restoring the agreement with the Bethe ansatz solution. It is quite possible that the same argument could work also in this case but since it involves also loop-corrections to fermion two-point functions we would need the string action to quartic order in fermions in order to check it. We are therefore not able to say definitively whether the heavy pole disappears or persists, at least for general $\phi$.

\subsubsection{Light modes}
Let us now turn to the two light coordinates $y_2$ and $y_3$. As for the heavy mode, we have three-vertex bubble and tadpole diagrams. Since the cubic Lagrangian has quadratic $y_2$ and $y_3$ terms we could, in principle, have non-zero tadpoles built out of three-vertices. However, due to a delicate cancellation between fermion and boson loops these diagrams do not contribute,
\[
\parbox[top][0.6in][c]{1in}{\fmfreuse{lollF}}+\qquad 
\parbox[top][0.6in][c]{1in}{\fmfreuse{lollB}}\qquad \qquad =0\]
and similarly for $y_3$. The remaining three-vertex bubble diagrams are, however, not zero and from (\ref{L3-full1}) we see that
\begin{align}
\mathcal{A}_{B}^2 & =\\ \nn 
&\parbox[top][0.8in][c]{1.5in}{\fmfreuse{bubble-y2-chi24}} +\quad 
\parbox[top][0.8in][c]{1.5in}{\fmfreuse{bubble-y2-chi13}}+ \quad\parbox[top][0.8in][c]{1in}{\fmfreuse{bubble-y2-y24}}, \\ \nn 
\mathcal{A}_{B}^3 & =\\ \nn 
&\parbox[top][0.8in][c]{1.5in}{\fmfreuse{bubble-y3-chi34}} +\quad 
\parbox[top][0.8in][c]{1.5in}{\fmfreuse{bubble-y3-chi12}}+ \quad\parbox[top][0.8in][c]{1in}{\fmfreuse{bubble-y3-y34}}
\end{align}
so in each loop there is always a light field with either a massless or heavy mode propagating. The actual form of the amplitude is rather involved and for this reason we will put $\phi=\pi/4$. What is more, since we have massless modes propagating in the loop we need to regularize the IR sector and this we do by introducing a small non-zero mass $m_4$ for $y_4$ and $\chi_\pm^{(4)}$. The sum of integrals is fairly involved and for the first coordinate, $y_2$, we find
\begin{align}
\nn &
\mathcal{A}^2_B\vert_{\phi=\frac{\pi}{4}}=-\frac{1}{256\pi^2}\int_0^1 dx\Big[\big((1+x)^2+8(-1+x^2)p_1^2\big)\,I^0_2[\frac{(1+x)^2}{4}]\\ \nn 
& +(-2+x)\big(-2+x+8x\,p_1^2\big)I^0_2[\frac{(x-2)^2}{4}] 
+2(1-x)\big(m_4+2(1-x)p_1^2\big)I^0_2[\frac{(x-1)^2+4 xm_4^2}{4}] \\ \nn 
&
-2x\big(x-m_4+2x\,p_1^2\big)I^0_2[\frac{x^2+4(1-x)m_4^2}{4}]\Big]+2\Big[8p_1^2 I^1_2[\frac{(x-1)^2+4x m_4^2}{4}] \\ \nn 
& -2\big(1-4p_1^2\big)I^1_2[\frac{x^2+4(1-x)m_4^2}{4}]+\big(1+8p_1^2\big)\big(I^1_2[\frac{(x-2)^2}{4}]+I^1_2[\frac{(x+1)^2}{4}]\big)
\Big]\,,
\end{align}
while the amplitude for $y_3$ is given by
\begin{align}
\nn &
\mathcal{A}^3_B\vert_{\phi=\frac{\pi}{4}}=\frac{-1}{256\pi^2}\int_0^1 dx\Big[
-3x^2\big(1+4p_1^2\big)I^0_2\big[\frac{x^2}{4}+(2-x)m_4^2\big] +2\big((1+x)^2-8(1-x^2)p_1^2\big)I^0_2\big[\frac{(1+x)^2}{4}\big]
\\ \nn 
& -(1-x)\big(-1+x-4m_4+12(-1+x)p_1^2\big)I^0_2\big[\frac{(x-1)^2}{4}+xm_4^2\big]
+2\big(1+16p_1^2\big)I^1_2\big[\frac{(1-x)^2+4xm_4^2}{4}\big] \\ \nn 
& +4\big(1+8p_1^2\big)I^1_2\big[\frac{(1+x)^2}{4}\big]-6I^1_2\big[\frac{x^2+4(1-x)m_4^2}{4}\big]\Big]\,.
\end{align}
The amplitudes are evaluated at $p_0=\sqrt{\frac{1}{4}+p_1^2}$ and we see that we have UV and possibly IR divergent terms. A quick evaluation gives
\bea 
\mathcal{A}^i_B\vert_{\phi=\frac{\pi}{4}}=-i\frac{\gamma+\log\pi-\frac{2}{\epsilon}}{4\pi}p_1^2-i\frac{\log 2}{16\pi}+i\frac{\log 2}{2\pi}p_1^2,\qquad i=2,3
\eea 
where it is gratifying to see that all IR divergent terms cancel within the amplitude. Since we are looking at the special case where $\phi=\pi/4$, it's of course expected that the amplitudes for $y_2$ and $y_3$ should be the same.

We now turn to the four-vertex tadpoles which are made out of the following diagrams
\begin{align*}
\mathcal{A}^i_T=\parbox[top][0.8in][c]{1in}{\fmfreuse{tadpole-y2ferm}} +
\parbox[top][0.8in][c]{1.5in}{\fmfreuse{tadpole-y2bos}}
\end{align*}
where $i=2,3$. In the fermion loop any of the $\chi^{(k)}_\pm$ fields can propagate while in the boson loop we only have $y_1,y_2$ and $y_3$ propagating. As for the bubble the actual amplitude is rather involved but simplifies in the $\phi=\pi/4$ case to
\begin{align}
& \mathcal{A}^i_T\vert_{\phi=\frac{\pi}{4}}= \\ \nn
&-\frac{1}{32\pi^2}\Big[2\big(1+4p_1^2\big)I^0_1\big[\frac{1}{4}\big]-\big(1+16p_1^2\big)I^0_1\big[1\big]
+m_4 I^0_1\big[m_4^2\big]  -4 I^1_1 \big[\frac{1}{4}\big]-2 I^1_1\big[ m_4^2\big]\Big],\qquad i=2,3\,.
\end{align}
Using the integral representations we find
\bea 
\mathcal{A}^i_T\vert_{\phi=\frac{\pi}{4}}=i\frac{\gamma+\log\pi-\frac{2}{\epsilon}}{4\pi}p_1^2-i\frac{\log\frac{\beta}{2}}{16\pi}+i\frac{\log 2 -2\log\beta}{2\pi}p_1^2\,,
\eea 
which is manifestly IR finite. Summing up the tadpole and bubble contributions gives
\bea \label{y2y3amp}
\mathcal{A}^i\vert_{\phi=\frac{\pi}{4}}=-i\frac{\log \beta}{16\pi}-i\frac{\log \frac{\beta}{2}}{\pi}\,p_1^2\qquad i=2,3\,.
\eea 
Here we note that for the AC ($\beta=2$) regulator, we obtain a finite mass renormalization. This is somewhat unexpected since supersymmetry should prevent any mass renormalization of the light modes. Perhaps this is an indication that $\beta=1$ is the regulator that preserves worldsheet supersymmetry. However, to make an explicit statement one should derive the full supersymmetric action with quartic fermions. It would be very interesting to see whether the fermionic coordinates $\chi_\pm^{(2)}$ and $\chi_\pm^{(3)}$ exhibit the same one-loop mass renormalization. This will also affect the question of whether the heavy mode disappears or not since it decays into the light fermions.

The $\log\frac{\beta}{2}$ term implies that we find a subleading correction to $h(\lambda)$ identical to the one found in the $AdS_4 / CFT_3$ case,
\bea \label{eq:calculated-h}
h(\lambda)\vert_{\phi=\frac{\pi}{4}}=\sqrt\frac{\lambda}{2}+\frac{\log \frac{\beta}{2}}{2\pi}+\ldots
\eea 
A similar term was found in \cite{Forini:2012bb} where folded string solutions were considered, however there the equivalent of $\beta=1$ was used. Naturally, for general three-sphere radii the subleading correction to $h(\lambda)$ will depend on $\phi$.

It is also interesting to look at the limit where one of the three-spheres is 'decompactified'. For $\phi=0,\pi/2$, the cubic vertices completely vanish and the tadpole is relatively simple. A quick calculation gives,
\bea 
\mathcal{A}^i\vert_{\phi=0,\frac{\pi}{2}}=0
\eea 
for both $y_2$ and $y_3$ irrespective of whether we take $\phi=0$ or $\phi=\frac{\pi}{2}$. As for the heavy mode this result is identical for both AC and WS regularization schemes. As we mentioned earlier this is expected from corresponding $AdS_5\times S^5$ and $AdS_2\times S^2\times T^6$ calculations. Furthermore, using coset model constructions, the exact solutions based on properties of $PSU(2,2|4)$ and $PSU(1,1|2)$ also support this. In these models all excitations come with the same mass and the notion of heavy composite modes are absent and hence we expect non ambiguous results. It would be interesting to reconcile this result with the corresponding $AdS_3\times S^3\times T^4$ computation of \cite{Forini:2012bb} which found a non-zero correction for $h(\lambda)$ (albeit different than the correction for the $\phi=\frac{\pi}{4}$ case).

\subsubsection{Massless mode}
We now turn to the massless mode, $y_4$. As for the light modes, the three-vertex tadpoles cancel between fermion and boson loops. The bubble diagrams, on the other hand, are not zero and the loops are made out of the two light fields,
\begin{align}
\mathcal{A}_{B}^4 & =
\parbox[top][0.8in][c]{1.5in}{\fmfreuse{bubble-y4-chi23}}+\quad
\parbox[top][0.8in][c]{1.5in}{\fmfreuse{bubble-y4-y23}} \qquad \textrm{where}  \quad i=2,3 \\ \nn 
& =-\frac{1}{32\pi^2}\sin^22\phi\int^1_0 dx\big(m_4^2+p_1^2\big)\Big[4\big((-1+x)x m_4^2-2\sin^4\phi\big) I^0_2\big[\sin^4\phi-(1-x)x m_4^2\big]\\ \nn 
&+4\big((-1+x)x m_4^2-2\cos^4\phi\big) I^0_2\big[\cos^4\phi-(1-x)x m_4^2\big]+4I^1_2\big[\cos^4\phi-(1-x)x m_4^2\big]\\ \nn 
&+4I^1_2\big[\sin^4\phi-(1-x)x m_4^2\big]\Big]
\\ \nn 
& = -i\frac{\gamma+\log\pi-\frac{2}{\epsilon}}{4\pi}p_1^2\sin^22\phi+i\frac{1-2\log\cos\phi\sin\phi}{4\pi}p_1^2\sin^22\phi
\end{align}
which is IR finite. For the four-vertex tadpoles, only the purely bosonic piece of $\mathcal{L}_4$ contributes,
\begin{align}
\mathcal{A}_{T}^4&  =
\parbox[top][0.8in][c]{1in}{\fmfreuse{tadpole-y4}}  = -\frac{1}{4\pi^2}\Big[\cos^2\phi\big(p_1^2+\cos 2\phi(m_4^2+p_1^2)\big)I^0_1\big[\cos^4\phi\big]\\ \nn 
&+ \sin^2\phi\big(p_1^2-\cos 2\phi(m_4^2+p_1^2)\big)I^0_1\big[\sin^4\phi\big]-\big(m_4^2+2p_1^2\big)I^0_1\big[1\big]\Big]\,. %\\ \nn 
%&= -i\frac{\gamma+\log\pi-\frac{2}{\epsilon}}{4\pi}p_1^2\sin^22\phi+\frac{i}{\pi}\big(\log\beta+2(\cos^4\log\cos\phi+\sin^4\phi\log\sin\phi)\big)p_1^2
\end{align}
Summing up the two contributions gives
\bea \label{eq:masslessgeneralFI}
&&
\mathcal{A}^4_{AC}=\frac{i\big(1-2\log \sin 2\phi\big)}{4\pi}\sin^2 2\phi\,p_1^2,\\ \nn 
&&
\mathcal{A}^4_{WS}=\frac{i}{4\pi}\sin^2 2\phi\,p_1^2-2i\frac{\cos^2\phi\log\cos\phi+\sin^2\phi\log\sin\phi}{\pi}\,p_1^2
\eea 
which in the equal radii case simplifies to
\bea \label{eq:y4equalradii}
\mathcal{A}^4\vert_{\phi=\frac{\pi}{4}}=\frac{i}{4\pi}\big(1-\log\frac{\beta}{2}\big)p_1^2\,.
\eea 
Compared to the other coordinates we see that we have a novel $p_1^2$ dependent term lingering around. This term is similar to the $AdS_4\times\mathbbm{CP}^3$ case where similar terms were found \cite{Abbott:2011xp}. However, here we would like to point out the following observation: If we use the identity
\bea \nn 
\frac{1}{2}\int d^2\ell \,\partial_\mu \big(\frac{\ell^\mu}{\ell^2-\Delta}\big)=I^0_1\big[\Delta\big]-I^1_2\big[\Delta\big]=-\frac{i}{2}\pi
\eea 
where we in the last line evaluated each integral explicitly, it becomes clear that the first term in (\ref{eq:y4equalradii}) originates from a surface term. Since the integrals in DREG are assumed to be invariant under shifts of the loop variable, one disregards surface terms of this kind. Hence one could, at least within the regularization scheme employed here, argue that they are unphysical. See \cite{Ferreira:2011cv} for a nice discussion. 

As we did for the light and heavy modes, let's also investigate what happens in the $\phi=0,\frac{\pi}{2}$ case. From (\ref{eq:masslessgeneralFI}) we find
\bea 
\mathcal{A}^4\vert_{\phi=0,\pi/2}=0
\eea 
for both AC and WS schemes.

\section{Summary}
The $AdS_3\times S^3\times S^3\times S^1$ and $AdS_3\times S^3\times T^4$ strings, related through the parameter $\phi$, are new examples of integrable sigma models. In this paper we proved classical integrability of the string up to quadratic order in fermions (without fixing kappa-symmetry) and computed one-loop corrections to the propagators for the bosonic coordinates. The string sigma model shares many features with both the $AdS_4\times \mathbbm{CP}^3$ and $AdS_5\times S^5$ strings and , in particular, we found that the one-loop contributions exhibit the same kind of regularization ambiguities as in $AdS_4 / CFT_3$. 

There are several issues that warrant further investigation. First, and perhaps foremost, one should compute the S-matrix \cite{Ahn:2008aa}. This would give direct information on how the massless modes interact with the massive ones. As of yet it is unclear how to incorporate these in the Bethe ansatz. At the level of the sigma model there is another pressing issue: One should sort out the regularization ambiguity once and for all. This is especially important since the ambiguity seems to be a rather generic feature for $AdS / CFT$ with heavy and light modes. 

Another interesting line of research would be to derive the full quartic Lagrangian. With the missing quartic interactions one could perform the corresponding one-loop computations also for the fermionic coordinates $\chi_\pm^{(i)}$. It would be especially interesting to see whether the one-loop mass renormalization for the light fermions are the same as for the light bosons. This would also allow for a more detailed  analysis of the compositeness of the heavy mode.

We hope to return to several of these questions in an upcoming paper. 

\section*{Acknowledgments}
It's a great pleasure to thank Michael Abbott, Dmitri Bykov, Horatiu Nastase, Nitin Rughoonauth, Christoph Sieg, Dima Sorokin and Kostya Zarembo for many illuminating discussions and useful comments.

P.S is supported by a postdoctoral grant from the Claude Leon Foundation. The research of L.W is supported in part by NSF grants PHY-0555575 and PHY-0906222. 

%-----------------------------APPENDIX-------------------------------

\begin{appendix}
\label{sec:appendix}
\section*{Appendix}

\section{The $AdS_3\times S^3\times S^3\times S^1$ superisometry algebra}
The superisometry group of $AdS_3\times S^3\times S^3\times S^1$ contains two copies of the exceptional supergroup $D(2,1;\alpha)$, where the parameter $\alpha=\cos^2\phi$ determines the relative radii of the two $S^3$ according to (\ref{eq:triangel-identity}). To describe the superisometry algebra we first need to describe the algebra of $D(2,1;\alpha)$.

\subsection{The superalgebra of $D(2,1;\alpha)$}
The $D(2,1;\alpha)$ superalgebra consists of 9 bosonic generators $S_{A'}$ ($A'=0,\ldots,8$) and 8 fermionic generators $Q_{\alpha'}$ ($\alpha'=1,\ldots,8$). The commutators of the bosonic generators are (see Appendix A of \cite{Babichenko:2009dk})
\begin{eqnarray}
[S_{A'},S_{B'}]=\varepsilon_{A'B'C'}S^{C'}\qquad (A',B',C'=0,\ldots,8)\,,
\end{eqnarray}
where $\varepsilon_{A'B'C'}$ is antisymmetric with non-zero components $\varepsilon_{012}=\varepsilon_{345}=\varepsilon_{678}=1$. The commutation relations involving the fermionic generators take the form
\begin{eqnarray}
[S_{A'},Q_{\alpha'}]&=&-(-i)^{A'}\frac{i}{2}Q_{\beta'}(\tilde\gamma_{A'})^{\beta'}{}_{\alpha'}
\nonumber\\
\{Q_{\alpha'},Q_{\beta'}\}&=&(\tilde C\tilde\gamma^a)_{\alpha'\beta'}\,S_a+i\cos^2\phi\,(\tilde C\tilde\gamma^{\hat a})_{\alpha'\beta'}\,S_{\hat a}+i\sin^2\phi\,(\tilde C\tilde\gamma^{a'})_{\alpha'\beta'}\,S_{a'}\,,
\end{eqnarray}
where $(-i)^a=-i$ and $(-i)^{\hat a}=1=(-i)^{a'}$ with $a=0,1,2$, $\hat a=3,4,5$ and $a'=6,7,8$. The $8\times8$ matrices $\tilde\gamma$ and $\tilde C$ can be chosen as follows
\begin{eqnarray} \label{eq:gammamatrices}
\tilde\gamma^a&=&\rho^a\otimes\mathbbm1\otimes\mathbbm1\,,\qquad\rho^a=(i\sigma^2,\sigma^1,\sigma^3)\,,\nonumber\\
\tilde\gamma^{\hat a}&=&\mathbbm1\otimes\rho^{\hat a}\otimes\mathbbm1\,,\qquad\rho^{\hat a}=(\sigma^1,\sigma^2,\sigma^3)\,,\nonumber\\
\tilde\gamma^{a'}&=&\mathbbm1\otimes\mathbbm1\otimes\rho^{a'}\,,\qquad\rho^{a'}=(\sigma^1,\sigma^2,\sigma^3)\,,\nonumber\\
\tilde C&=&\sigma^2\otimes\sigma^2\otimes\sigma^2\,.
\end{eqnarray}
We are now ready to write down the full superisometry algebra.

\subsection{The $D(2,1;\alpha)\times D(2,1;\alpha)\times U(1)$ superisometry algebra}
The superisometry algebra of $AdS_3\times S^3\times S^3\times S^1$ involves two copies of $D(2,1;\alpha)$. Denoting the generators of the second $D(2,1;\alpha)$ with hats we form the following combinations of the bosonic generators
\begin{eqnarray}
P_A&=&\left(-\frac{(-1)^{A'}}{R_{A'}}(S-\hat S)_{A'}\,,\,P_9\right)\qquad \left(R_{A'}=(R,\,R_+,\,R_-)\quad\mbox{for}\quad A'=(a,\,\hat a,\,a')\right)\,,
\nonumber\\
M_{AB}&=&M_{A'B'}=(-1)^{A'}\varepsilon_{A'B'C'}(S+\hat S)^{C'}\,,
\end{eqnarray}
where we have added also the $U(1)$-generator $P_9$ corresponding to translations along the $S^1$. Recall that the radii of the factors in $AdS_3\times S^3\times S^3$ are respectively $R$, $R_+=\frac{R}{\cos\phi}$ and $R_-=\frac{R}{\sin\phi}$ where $\cos^2\phi=\alpha$. Using the form of the algebra of $D(2,1;\alpha)$ in the previous section it is easy to show that the commutation relations of these new generators take the form
\begin{eqnarray}
[P_A,P_B]=-\frac{1}{2}R_{AB}{}^{CD}M_{CD}\,,\qquad[M_{AB},P_C]=\eta_{AC}P_B-\eta_{BC}P_A\,,
\nonumber\\
{}[M_{AB},M_{CD}]=\eta_{AC}M_{BD}+\eta_{BD}M_{AC}-\eta_{BC}M_{AD}-\eta_{AD}M_{BC}\,,
\label{eq:bosonic-algebra}
\end{eqnarray}
where the non-zero components of the Riemann curvature tensor $R_{AB}{}^{CD}$ are
\begin{equation}
R_{ab}{}^{cd}=\frac{2}{R^2}\delta_{[a}^c\delta_{b]}^d\,,\qquad R_{\hat a\hat b}{}^{\hat c\hat d}=-\frac{2}{R_+^2}\delta_{[\hat a}^{\hat c}\delta_{\hat b]}^{\hat d}\,,\qquad R_{a'b'}{}^{c'd'}=-\frac{2}{R_-^2}\delta_{[a'}^{c'}\delta_{b']}^{d'}\,.
\end{equation}
For the fermionic generators we define new generators $Q_{\alpha'i}$ ($i=1,2$) as the combinations
\begin{equation}
Q_1=\sqrt{\frac{2i}{R}}\,(Q-i\hat Q)\,,\qquad Q_2=\sqrt{\frac{2i}{R}}\,(\hat Q-iQ)\,.
\end{equation}
Using the form of the $D(2,1;\alpha)$ algebra in the previous section one finds the new commutation relations (for readability we suppress the $\alpha'$-indices)
\begin{eqnarray}
\label{eq:Qcomm1}
[P_{A'},Q_i]&=&\frac{i^{A'+1}}{2R_{A'}}(Q_j\tilde\gamma_{A'})(\sigma^2)^j{}_i\,,\qquad[M_{A'B'},Q_i]=-\frac{1}{2}(Q_i\tilde\gamma_{A'B'})\,,
\nonumber\\
\{Q_i,Q_j\}&=&2i\sigma^3_{ij}(\tilde C\tilde\gamma^a)\,P_a+2\cos\phi\,\sigma^3_{ij}(\tilde C\tilde\gamma^{\hat a})\,P_{\hat a}+2\sin\phi\,\sigma^3_{ij}(\tilde C\tilde\gamma^{a'})\,P_{a'}
\nonumber\\
&&{}
-\frac{1}{R}\sigma^1_{ij}(\tilde C\tilde\gamma^{ab})\,M_{ab}
+\frac{\cos^2\phi}{R}\sigma^1_{ij}(\tilde C\tilde\gamma^{\hat a\hat b})\,M_{\hat a\hat b}
+\frac{\sin^2\phi}{R}\sigma^1_{ij}(\tilde C\tilde\gamma^{a'b'})\,M_{a'b'}\,.
\end{eqnarray}
We want to write these commutators in terms of standard $32\times32$ gamma-matrices. To this end we use the gamma-matrix realization in Appendix A of \cite{Babichenko:2009dk} which reads, in our notation,
\begin{eqnarray}
\Gamma^a=\sigma^1\otimes\sigma^2\otimes\tilde\gamma^a\,,\quad
\Gamma^{\hat a}=\sigma^1\otimes\sigma^1\otimes\tilde\gamma^{\hat a}\,,\quad
\Gamma^{a'}=\sigma^1\otimes\sigma^3\otimes\tilde\gamma^{a'}\,,\quad
\Gamma^9=-\sigma^2\otimes\mathbbm1\otimes\mathbbm1\,,
\end{eqnarray}
with the charge-conjugation matrix $\mathcal C=i\sigma^2\otimes\sigma^2\otimes\tilde C$. This gamma-matrix realization is convenient since the projection matrix which singles out the supersymmetries defined in (\ref{eq:FP}) simply becomes
\begin{equation}
\mathcal P=\frac{1}{2}(1+\cos\phi\,\Gamma^{012345}+\sin\phi\,\Gamma^{012678})=
\mathbbm1\otimes
\frac{1}{2}(
\mathbbm1
+\cos\phi\,\sigma^3
-\sin\phi\,\sigma^1
)
\otimes\mathbbm1_{8\times8}\,.
\end{equation}
Therefore when projected with $\mathcal P$ a $32$-component spinor index $\alpha=1,\ldots,32$ reduces to a 16-component index $(\alpha'i)$, $\alpha'=1,\ldots8$ and $i=1,2$ where the $i$ index refers to the first, $2\times2$, factor and the $\alpha'$ index to the last, $8\times8$, factor. We therefore find
\begin{eqnarray}
&&(\mathcal P\Gamma_*\Gamma^a\mathcal P)^{\alpha'i}{}_{\beta'j}=i\,(\sigma^2)^i{}_j(\tilde\gamma^a)^{\alpha'}{}_{\beta'}\,,\qquad
(\mathcal P\Gamma_*\Gamma^{\hat a}\mathcal P)^{\alpha'i}{}_{\beta'j}=\cos\phi\,(\sigma^2)^i{}_j(\tilde\gamma^{\hat a})^{\alpha'}{}_{\beta'}\,,\qquad
\nonumber\\
&&(\mathcal P\Gamma_*\Gamma^{a'}\mathcal P)^{\alpha'i}{}_{\beta'j}=\sin\phi\,(\sigma^2)^i{}_j(\tilde\gamma^{a'})^{\alpha'}{}_{\beta'}\,,\qquad
(\mathcal P\Gamma_*\Gamma^9\mathcal P)^{\alpha'i}{}_{\beta'j}=0\,,
\end{eqnarray}
where $\Gamma_*$ is defined in (\ref{eq:FP}) and is equal to $i\Gamma^{012}\Gamma^9$. We also get
\begin{eqnarray}
&&(\mathcal C\bar{\mathcal P}\Gamma^a\mathcal P)_{\alpha'i\beta'j}=\sigma^3_{ij}(\tilde C\tilde\gamma^a)_{\alpha'\beta'}\,,\qquad
(\mathcal C\bar{\mathcal P}\Gamma^{\hat a}\mathcal P)_{\alpha'i\beta'j}=-i\cos\phi\,\sigma^3_{ij}(\tilde C\tilde\gamma^{\hat a})_{\alpha'\beta'}\,,\qquad
\nonumber\\
&&(\mathcal C\bar{\mathcal P}\Gamma^{a'}\mathcal P)_{\alpha'i\beta'j}=-i\sin\phi\,\sigma^3_{ij}(\tilde C\tilde\gamma^{a'})_{\alpha'\beta'}\,,\qquad
(\mathcal C\bar{\mathcal P}\Gamma^9\mathcal P)_{\alpha'i\beta'j}=0\,,\qquad
\end{eqnarray}
where $\bar{\mathcal P}=-\mathcal C\mathcal P^T\mathcal C=(1-\mathcal P)$. Finally we have
\begin{eqnarray}
(\mathcal C\bar{\mathcal P}\Gamma_{AB}\Gamma_*\mathcal P)_{\alpha'i\beta'j}=-\sigma^1_{ij}(\tilde C\tilde\gamma_{AB})_{\alpha'\beta'}\,.
\end{eqnarray}
Using these relations in (\ref{eq:Qcomm1}) and replacing $Q_{\alpha'i}\rightarrow Q_\alpha=(Q\mathcal P)_\alpha$, the algebra involving $Q$ takes the form
\begin{eqnarray}
[P_A,Q]=\frac{i}{2R}Q\Gamma_*\Gamma_A\mathcal P\,,\qquad[M_{AB},Q]=-\frac{1}{2}Q\Gamma_{AB}\mathcal P
\nonumber\\
\{Q,Q\}=2i\mathcal C\bar{\mathcal P}\Gamma^A\mathcal P\,P_A
+\frac{R}{2}\mathcal C\bar{\mathcal P}\Gamma^{AB}\Gamma_*\mathcal P\,R_{AB}{}^{CD}M_{CD}\,,
\end{eqnarray}
where we have suppressed the (32-component) spinor indices. This, together with (\ref{eq:bosonic-algebra}), is the form of the superisometry algebra which we have found useful for discussing integrability. The superisometry algebras of $AdS_4\times\mathbbm{CP}^3$ and $AdS_2\times S^2\times T^6$ can also be cast into this form \cite{Sorokin:2010wn,Sorokin:2011rr}.

\section{Derivation of eq. (\ref{eq:deltaedual})}
The kappa-variation of $*e^A$ is given by
\begin{equation}
\delta_\kappa(*e^A)=
*\delta_\kappa(e^A)
+d\xi^k\,\frac{\varepsilon_{ki}}{-h}\,\delta_\kappa(\sqrt{-h}h^{ij})e_j{}^A\,.
\end{equation}
Using the form of the kappa-variation of the worldsheet metric given in (\ref{eq:deltahij}) and the fact that we are working on-shell, so that we can replace $H^{ij}\rightarrow h^{ij}=g^{ij}$, the second term becomes
\begin{eqnarray}
\lefteqn{\frac{2i}{\sqrt{-g}}d\xi^m\,\varepsilon_{mi}
\left(g^{ij}g^{kl}-2g^{k(i}g^{j)l}\right)
e_j{}^Ae_k{}^B\,\delta_\kappa\Theta\Gamma_B{\mathcal D}_l\Theta}
\nonumber\\
&=&
\frac{2i}{\sqrt{-g}}d\xi^m\,\varepsilon_{mi}
\left(
2g^{i[j}g^{l]k}
-g^{ik}g^{jl}
\right)
e_j{}^Ae_k{}^B\,\delta_\kappa\Theta\Gamma_B{\mathcal D}_l\Theta
\nonumber\\
&=&
-2i\left(
*e^B\,g^{ij}
+e^B\,\frac{\varepsilon^{ij}}{\sqrt{-g}}
\right)
e_i{}^A\,\delta_\kappa\Theta\Gamma_B{\mathcal D}_j\Theta\,.
\label{eq:deltaedual1}
\end{eqnarray}
We wish to show that this is equal to
\begin{equation}
-2i\delta_\kappa\Theta\Gamma^A*\mathcal D\Theta
-2i\delta_\kappa\Theta\Gamma^A\Gamma_{11}\mathcal D\Theta\,.
\end{equation}
Due to the projection involved in $\delta_\kappa\Theta$, see (\ref{eq:deltakappa}), we can write this as
\begin{equation}
-i\delta_\kappa\Theta(1+\Gamma)\Gamma^A*\mathcal D\Theta
-i\delta_\kappa\Theta(1+\Gamma)\Gamma^A\Gamma_{11}\mathcal D\Theta\,.
\end{equation}
We now use the fact that
\begin{eqnarray}
\lefteqn{\frac{i}{2\sqrt{-g}}\,\varepsilon^{ij}e_i{}^Be_j{}^C\,\delta_\kappa\Theta
\left(
\Gamma^A{}_{BC}\Gamma_{11}*\mathcal D\Theta
+\Gamma^A{}_{BC}\mathcal D\Theta
\right)}
\nonumber\\
&=&
-\frac{id\xi^k}{\sqrt{-g}}\,\varepsilon^{ij}e_k{}^Be_i{}^C\,\delta_\kappa\Theta
\left(
\Gamma^A{}_{BC}\Gamma_{11}(*\mathcal D)_j\Theta
+\Gamma^A{}_{BC}\mathcal D_j\Theta
\right)
\nonumber\\
&=&
i\delta_\kappa\Theta\Gamma^A\Gamma_{11}\mathcal D\Theta
+i\delta_\kappa\Theta\Gamma^A*\mathcal D\Theta
-\frac{id\xi^k}{\sqrt{-g}}\,\varepsilon^{ij}e_k{}^Be_i{}^A\,\delta_\kappa\Theta
\left(
\Gamma_B\Gamma_{11}(*\mathcal D)_j\Theta
+\Gamma_B\mathcal D_j\Theta
\right)\,,
\end{eqnarray}
where we've made use of the equation of motion in the last step. Using this fact and the form of $\Gamma$ in (\ref{eq:deltakappa}) we find
\begin{eqnarray}
\lefteqn{
-i\delta_\kappa\Theta(1+\Gamma)\Gamma^A*\mathcal D\Theta
-i\delta_\kappa\Theta(1+\Gamma)\Gamma^A\Gamma_{11}\mathcal D\Theta
}
\nonumber\\
&=&
-ie^B\,\frac{\varepsilon^{ij}}{\sqrt{-g}}e_i{}^A\,\delta_\kappa\Theta\Gamma_B\mathcal D_j\Theta
-\frac{id\xi^k}{\sqrt{-g}}\,\varepsilon^{ij}e_k{}^Be_i{}^A\,\delta_\kappa\Theta\Gamma_B\Gamma_{11}(*\mathcal D)_j\Theta
\nonumber\\
&&{}
+\frac{i}{\sqrt{-g}}\,\varepsilon^{ij}e_i{}^Be_j{}^A\,\delta_\kappa\Theta\Gamma_B\Gamma_{11}*\mathcal D\Theta
+\frac{i}{\sqrt{-g}}\,\varepsilon^{ij}e_i{}^Be_j{}^A\,\delta_\kappa\Theta\Gamma_B\mathcal D\Theta
\nonumber\\
&=&
-\frac{2id\xi^k}{\sqrt{-g}}\,\varepsilon^{ij}e_k{}^Be_i{}^A\,\delta_\kappa\Theta\Gamma_B\Gamma_{11}(*\mathcal D)_j\Theta
-ie^B\,\frac{\varepsilon^{ij}}{\sqrt{-g}}e_i{}^A\,\delta_\kappa\Theta\Gamma_B\mathcal D_j\Theta
\nonumber\\
&&{}
+\frac{i}{\sqrt{-g}}\,\varepsilon^{ij}e_i{}^Be_j{}^A\,\delta_\kappa\Theta\Gamma_B\mathcal D\Theta
+\frac{id\xi^k}{\sqrt{-g}}\,e_k{}^A\varepsilon^{ij}e_i{}^B\,\delta_\kappa\Theta\Gamma_B\Gamma_{11}(*\mathcal D)_j\Theta\,.
\end{eqnarray}
The first term can be rewritten as follows
\begin{eqnarray}
\lefteqn{
-\frac{2id\xi^k}{\sqrt{-g}}\,\varepsilon^{ij}e_k{}^Be_i{}^A\,\delta_\kappa\Theta\Gamma_B\Gamma_{11}(*\mathcal D)_j\Theta
=
-\frac{2id\xi^k}{\sqrt{-g}}\,\varepsilon^{ij}e_k{}^Be_i{}^A\,\delta_\kappa\Theta\Gamma\Gamma_B\Gamma_{11}(*\mathcal D)_j\Theta
}
\nonumber\\
&=&
\frac{2id\xi^k}{-g}\,\varepsilon^{ij}g_{km}e_i{}^A\,\varepsilon^{lm}e_l{}^C\,\delta_\kappa\Theta\Gamma_C(*\mathcal D)_j\Theta
=
-2i*e^B\,g^{ij}e_i{}^A\,\delta_\kappa\Theta\Gamma_B\mathcal D_j\Theta\,.
\end{eqnarray}
Therefore we have
\begin{eqnarray}
\lefteqn{
-i\delta_\kappa\Theta(1+\Gamma)\Gamma^A*\mathcal D\Theta
-i\delta_\kappa\Theta(1+\Gamma)\Gamma^A\Gamma_{11}\mathcal D\Theta
}
\nonumber\\
&=&
-i\left(
2*e^B\,g^{ij}
+e^B\,\frac{\varepsilon^{ij}}{\sqrt{-g}}
\right)
e_i{}^A\,\delta_\kappa\Theta\Gamma_B\mathcal D_j\Theta
+\frac{i}{\sqrt{-g}}\,\varepsilon^{ij}e_i{}^Be_j{}^A\,\delta_\kappa\Theta\Gamma_B\mathcal D\Theta
\nonumber\\
&&{}
+\frac{id\xi^k}{\sqrt{-g}}\,e_k{}^A\varepsilon^{ij}e_i{}^B\,\delta_\kappa\Theta\Gamma_B\Gamma_{11}(*\mathcal D)_j\Theta
\nonumber\\
&=&
-i\left(
2*e^B\,g^{ij}
+e^B\,\frac{\varepsilon^{ij}}{\sqrt{-g}}
\right)
e_i{}^A\,\delta_\kappa\Theta\Gamma_B\mathcal D_j\Theta
+\frac{id\xi^k}{\sqrt{-g}}\,\varepsilon^{ij}e_i{}^Be_j{}^A\,\delta_\kappa\Theta\Gamma_B\mathcal D_k\Theta
\nonumber\\
&&{}
-\frac{id\xi^k}{\sqrt{-g}}\,\varepsilon^{ij}e_k{}^Ae_i{}^B\,\delta_\kappa\Theta\Gamma_B\mathcal D_j\Theta
\nonumber\\
&=&
-2i\left(
*e^B\,g^{ij}
+e^B\,\frac{\varepsilon^{ij}}{\sqrt{-g}}
\right)
e_i{}^A\,\delta_\kappa\Theta\Gamma_B\mathcal D_j\Theta\,,
\end{eqnarray}
where we've used the equations of motion in the next-to-last step. This agrees with (\ref{eq:deltaedual1}) which completes the proof of (\ref{eq:deltaedual}).

\section{Relevant piece of quartic Lagrangian}
Here we collect the piece of the quartic Lagrangian that is needed for our one-loop computations\footnote{To keep the expression as compact as possible we here denote $\partial_+$ with dot and $\partial_-$ with prime.}
\begin{align}
\label{Lbf}
\mathcal{L}^4_{BF} &= \frac{i}{4}\Big(\dot{\chi}^i_+\bar{\chi}^i_+ + (\chi^i_-)' \bar{\chi}^i_-\Big)\ |y_1|^2 \\ \nn 
&\quad - \frac{i}{4}\cos^4\phi\ \left[\dot{\chi}^i_+\bar{\chi}^i_+ + (\chi^i_-)' \bar{\chi}^i_- - 4i\ \sin^2\phi\ \left(\chi^2_- \bar{\chi}^2_+ - \chi^3_- \bar{\chi}^3_+\right)\right]|y_2|^2\\ \nonumber
&\quad - \frac{i}{4}\sin^4\phi\ \left[\dot{\chi}^i_+ \bar{\chi}^i_+ + (\chi^i_-)' \bar{\chi}^i_- + 4i\ \cos^2\phi\ \left(\chi^2_- \bar{\chi}^2_+ - \chi^3_- \bar{\chi}^3_+\right)\right]|y_3|^2\\ \nonumber
&\quad - \frac{1}{2}\left(\chi^1_-\bar{\chi}^1_+ + \cos^2\phi\ \chi^2_-\bar{\chi}^2_+ + \sin^2\phi\ \chi^3_+\bar{\chi}^3_-\right)\dot{\bar{y}}_1 y'_1\\ \nonumber
&\quad - \frac{i}{4}\left[\left(\chi^1_-\bar{\chi}^1_- + \chi^2_-\bar{\chi}^2_- - \chi^3_-\bar{\chi}^3_- - \chi^4_-\bar{\chi}^4_-\right)-\left(\chi^1_+\bar{\chi}^1_+ + \chi^2_+\bar{\chi}^2_+ - \chi^3_+\bar{\chi}^3_+ - \chi^4_+\bar{\chi}^4_+\right)\right]y_1(\dot{\bar{y}}_1 - \bar{y}'_1)\\ \nonumber
&\quad + \frac{1}{2}\left(\cos^2\phi\ \chi^1_+\bar{\chi}^1_- +  \chi^2_+\bar{\chi}^2_- + \sin^2\phi\ \chi^4_+\bar{\chi}^4_-\right)\ \dot{\bar{y}}_2 y'_2 -\frac{i}{4}\cos^2\phi\,\chi^i_-\bar\chi^i_-\,y_2\big(\dot{\bar{y}}_2-\cos^2\phi\bar{y}_2'\big)
\\ \nonumber
&\quad -\frac{i}{4}\cos^2\phi\,\chi^i_+\bar\chi^i_+\,y_2\big(\bar{y}_2'-\cos^2\phi\dot{\bar{y}}_2\big)
+\frac{1}{2}\left(\sin^2\phi\chi^1_+\bar{\chi}^1_- + \chi^3_-\bar{\chi}^3_+ + \cos^2\phi\chi^4_+\bar{\chi}^4_-\right)\dot{\bar{y}}_3 y'_3\\ \nonumber 
&\quad - \frac{i}{4}\sin^2\phi
\left(\chi^1_-\bar{\chi}^1_- - \chi^2_-\bar{\chi}^2_- - \chi^3_-\bar{\chi}^3_- + \chi^4_-\bar{\chi}^4_-\right)\,y_3(\dot{\bar{y}}_3-\sin^2\phi\ \bar{y}'_3)  \\ \nn 
& \quad -\frac{i}{4}\sin^2\phi\left(\chi^1_+\bar{\chi}^1_+ - \chi^2_+\bar{\chi}^2_+ - \chi^3_+\bar{\chi}^3_+ + \chi^4_+\bar{\chi}^4_+\right)\,y_3(\bar{y}'_3-\sin^2\phi\ \dot{\bar{y}}_3)
 \\ \nn 
&\quad + \frac{1}{2}\Big(\sin^2\phi\ \chi^2_-\bar{\chi}^2_+ +\cos^2\phi\ \chi^3_+\bar{\chi}^3_- + \chi^4_-\bar{\chi}^4_+\Big)\dot{y}_4 y'_4 \\ \nn 
&\quad + h.c. +\ldots\,,
\end{align}
where the ellipses denote parts not relevant for our computations.

\end{appendix}

\end{document}